\documentclass[journal=jacsat,manuscript=article]{achemso}
\usepackage[version=3]{mhchem} 
\usepackage{achemso}
\usepackage{graphicx} 
\usepackage{dcolumn} 
\usepackage{bm} 
\usepackage[utf8]{inputenc}
\usepackage[T1]{fontenc}  
\usepackage{hyperref} 
\usepackage{enumerate} 
\usepackage{float}
\usepackage{physics,color}
\usepackage{amsmath} 
\newcommand{\mycomment}[1]{\textcolor{black}{#1}}

\SectionNumbersOn
\author{Xuezhi Bian}
\affiliation{Department of Chemistry, University of Pennsylvania, Philadelphia, Pennsylvania 19104, USA}
\author{Yanze Wu}
\affiliation{Department of Chemistry, University of Pennsylvania, Philadelphia, Pennsylvania 19104, USA}
\author{Hung-Hsuan Teh}
\affiliation{Department of Chemistry, University of Pennsylvania, Philadelphia, Pennsylvania 19104, USA}
\author{Joseph E. Subotnik} 
\affiliation{Department of Chemistry, University of Pennsylvania, Philadelphia, Pennsylvania 19104, USA}
\title{Incorporating Berry Force Effects into The Fewest Switches Surface Hopping Algorithm: Intersystem Crossing and The Case of Electronic Degeneracy}
\email{subotnik@sas.upenn.edu}

\begin{document}

\begin{abstract}
We present a preliminary surface-hopping approach for modeling intersystem crossing (ISC) dynamics between four electronic states: one singlet and one (triply degenerate) triplet. In order to incorporate all Berry force effects, the algorithm requires that, when moving along an adiabatic surface associated with the triplet manifold, \mycomment{one must also keep track of a quasi-diabatic index (akin to a ``$m_s$'' quantum number) for each trajectory. For a simple model problem, we find that a great deal of new physics can be captured by our algorithm, setting the stage for larger, more realistic (or perhaps even {\em ab initio}) simulations in the future.}
\end{abstract}

\section{Introduction}   
Tully's fewest switches surface hopping (FSSH)  \cite{Tully1990} algorithm is one of the most 
successful nonadiabatic dynamics methods on account of its reasonable accuracy and extremely low computational cost  \cite{Tully1998, Barbatti2011, Mai2018, Nelson2020, Kapral2016}. 
Provided the nuclear motion is effectively classical,   one can invoke surface hopping to simulate a wide range of nonadiabatic problems, especially if one utilizes recent improvements to the algorithm accounting for decoherence  \cite{Bittner1995, Jasper2005, Subotnik2011,Subotnik2016}, velocity reversal and frustrated hops \cite{Jasper2001, Jasper2003}. In recent years,  surface hopping has been widely used to simulate many processes such as photo-excitation  \cite{Schwartz1994, Nelson2011,Zaari2015,Chakraborty2020}, electron transfer  \cite{Landry2014,Atkins2017} and scattering   \cite{Shenvi2009,Golibrzuch2014}.

Interestingly, however, as has been demonstrated recently  (see Ref. \citenum{Miao2019}), Tully's algorithm can fail dramatically if spin degrees of freedom enter into the nonadiabatic motion.  One particular case is when one considers Hamiltonians with spin-orbit interactions and the 
system has an odd number of electrons. In such a case, the electronic Hamiltonian becomes complex-valued  \cite{Mead1979} and the nuclei will experience an effective magnetic force (known as Berry's geometric phase force  \cite{Berry1993})  when  moving along an adiabatic surface.  
Recently, our research group has argued (with some success)  \cite{Miao2020,Wu2021}  that one can extend FSSH to complex-valued problems with reasonable accuracy by 
(i) explicitly including Berry force in the nuclear equation of motion and (ii) finding an optimal momentum rescaling direction. 
Benchmarking of such an extended surface hopping algorithm is  presently ongoing. 

Now, all of the developments in Ref. \citenum{Miao2019} are predicated on the idea of a two state Hamiltonian (designed to represent a simple radical system).  As such, one might wonder: what is the nature
of Berry force when there are more than two electronic states?
This problem can be naturally addressed by considering a singlet-triplet crossing, a minimal model of which
consists of four states.
For such a model system, one can show analytically that
the Hamiltonian can be made completely real-valued--and so one might hypothesize that there is no Berry force \mycomment{(see Eq. \ref{eq:bf} below)}. 
Nevertheless, model calculations  \cite{Bian2021} clearly show that \mycomment{strong pseudomagnetic forces can emerge for a real-valued singlet-triplet crossing}, strongly suggesting that
Berry forces arise from the presence of electronic degeneracy
(and not just from the presence of a complex-valued derivative coupling).  

Finally, we cannot emphasize enough that 
the concept of chiral induced spin selectivity (CISS) as mediated by intersystem crossing and spin orbit coupling (SOC) has become one of 
the hottest research areas today in physical chemistry  \cite{Naaman2019, Naaman2015, Naaman2012}. 
Moreover, several model (computational) studies have shown that the Berry (geometric) magnetic force {\em can} lead to strong spin selectivity  \cite{Wu2020, Wu2021:CI}. For this reason, there is today 
a strong motivation to develop new 
mixed quantum classical dynamics models capable of treating electronic spin.  

With this background in mind, we will present here a preliminary extension of FSSH that incorporates Berry force effects for Hamiltonians with degenerate manifolds of electronic states. For a  singlet-triplet crossing model, we will show that a modified FSSH can capture  spin-dependent branching ratios as well as
nuclear momenta fairly accurately. The paper is structured as follows: In Sec. \ref{sec:Methods} we introduce our model Hamiltonian, we propose the concept of triplet
``quasi-diabats'' as a means of labeling Berry forces in the presence of many degenerate (or nearly degenerate) electronic states, and we outline our present FSSH algorithm.  In Sec. \ref{sec:Results}, we present simulation results. In Sec. \ref{sec:Discussion}, we 
discuss the limitations of our approach and highlight open questions.  

A word about notation is now appropriate.\mycomment{Below, we will write all nuclear vectors with an arrow above them; bold face indicates an electronic operator or vector.} Roman letters $\{i,j,k\}$ index adiabatic electronic states, $\{a,b,c\}$ index quasi-diabatic states, the Greek letter $\lambda$ indexes the active adiabat, $\mu$ indexes the active quasi-diabat, and the  Greek letters $\{\alpha, \beta, \gamma\}$ index nuclear coordinates. The only exception to the above rules is that, for  the specific Hamiltonian in Eq. \ref{eq:4stateHamiltonian}, we will label  $x,y$  as the specific nuclear coordinates.
 
\section{\label{sec:Methods}Theory and Methods}
\subsection{From a two-state crossing model to a singlet-triplet crossing model}

As a means of motivating the four-state model below, let us begin by reviewing the standard two-state model that exemplifies a Berry force. For a system with a single unpaired electron with spin, the very simplest (complex-valued) Hamiltonian  has the form:

\begin{equation}\label{eq:2stateHamiltonian}
    H = A\begin{pmatrix} \cos\theta  &   \sin\theta  e^{i\phi}     \\
       {\sin\theta}   e^{-i\phi} & -\cos\theta \\  
    \end{pmatrix} 
\end{equation}

We define the functions $\theta(x)$ and $\phi(y)$ as:
\begin{equation} \label{eq:thetaphi}
\begin{aligned}
    \theta &\equiv \frac {\pi} 2 ({erf}(Bx) + 1)  \\
    \phi &\equiv Wy  \\        
\end{aligned}
\end{equation} 

Here $A$, $B$ and $W$ are constants 
and $x$ and $y$ are nuclear coordinates. According to Eqs. 
\ref{eq:2stateHamiltonian}-\ref{eq:thetaphi}, we imagine two diabats crossing in the $x$-direction, while the phase of their coupling is modulated in the $y$-direction.
Note that, while the diabats are flat in the $y$-direction, the adiabats are  flat in all directions ($x,y$). 
\mycomment{This model is not physical but rather has been chosen on purpose to make a point: if a trajectory is moving along a flat adiabat and veers away from a straight line, one must conclude that Berry forces are at play}
In the limit of slow nuclear motion, Berry showed that nuclei follow not only the static adiabatic force, but also an extra "magnetic" force\cite{Berry1993}:
\begin{equation} \label{eq:bf}  {\vec  F}_j^{Berry} =  \stackrel{\leftrightarrow}{ \Omega}_j  \cdot {\vec v} =  2\hbar {\rm Im}   \sum_{k\neq j} \left[ {\vec d}_{jk} (\vec  v \cdot \vec  d_{kj}) \right]  \end{equation}
where the Berry curvature is\cite{Mead1979,Sakurai1995,Shankar2012}
\begin{equation} \label{eq:Berrycurvature}\stackrel{\leftrightarrow}{ \Omega}_{j} = i \vec \nabla \times {\vec d}_{jj}\end{equation} 
or in index form:
\begin{equation} \label{eq:Berrycurvatureindex} { \Omega}_{j}^{\alpha\beta} = i \left( \nabla_\alpha d_{jj}^\beta - \nabla_\beta { d}_{jj}^\alpha\right) = i\sum_k (d_{jk}^\alpha d_{kj}^\beta -d_{jk}^\beta d_{kj}^\alpha) \end{equation} 
The derivative coupling between adiabatic states $j$ and ${k}$ is $\vec {d}_{jk} =\bra{\psi_k} \vec {\nabla} \ket{\psi_j}$, or in index form, ${d}_{jk}^\alpha =\bra{\psi_k}  {\nabla}_\alpha \ket{\psi_j} $. 
\mycomment{Note that, within a two state model, Eq. \ref{eq:bf} is zero if the electronic Hamiltonian is real-valued. Furthermore,  $\vec {F}^{Berry}_j \cdot {\vec v} = 0$; in other words, the Berry pseudomagnetic force cannot do any work.}

As a result of Eq. \ref{eq:bf}, for the two-state model in Eq. \ref{eq:2stateHamiltonian}, a
trajectory incident on the upper adiabat from $x = -\infty$ and exiting on the same adiabat towards $x = \infty$  
will accumulate a momentum shift in the $y$-direction of magnitude of $W$ (or $-W$ if it is on the lower adiabat). \cite{footnote1}
In other words, a proper adiabatic calculation with a Berry force shows  \cite{Miao2019} that:
\begin{equation} \label{eq:integral} \Delta p^y_1 = \int_{t=0}^{t=\infty} { F^{Berry,y}_{j}} dt = W \end{equation}
This semiclassical prediction can easily be verified by fully quantum calculations. Recently, Ref. \citenum{Wu2021} proposed an FSSH algorithm that  incorporates the Berry force in Eq. \ref{eq:bf} and so far the algorithm has been reasonably successful for a few two-state model problems; in particular, the algorithm has been able to recover both branching ratios and asymptotic nuclear momenta. 

Now, in order to describe a singlet-triplet crossing in an analogous manner, one would like to extend the two-state algorithm in Ref. \citenum{Wu2021} to a four-state model. To that end, the natural analogue of Eq. \ref{eq:2stateHamiltonian} is as follows:

\begin{equation}\label{eq:4stateHamiltonian} \bm H = A\begin{pmatrix} \cos\theta  &  \frac {\sin\theta} {\sqrt{3}}   &  \frac {\sin\theta  e^{i\phi}} {\sqrt{3}}    &\frac { \sin\theta e^{-i\phi}} {\sqrt{3}}   \\
    \frac {\sin\theta} {\sqrt{3}}   & -\cos\theta  & 0 &0 \\ 
    \frac {\sin\theta e^{-i\phi}} {\sqrt{3}} & 0 & -\cos\theta & 0 \\
    \frac {\sin\theta  e^{i\phi}}  {\sqrt{3}}  & 0 & 0 & -\cos\theta
\end{pmatrix} \end{equation}
In Eq. \ref{eq:4stateHamiltonian}, the Hamiltonian is represented in the  $\{\ket{S}$, $\ket{T_0}$, $\ket{T_1}$  
and $\ket{T_{-1}}\}$   spin-diabatic basis. The diabatic and adiabatic surfaces are plotted in Fig. \ref{fig:geometry}.
Within this model, three triplets cross a singlet state, and the diabatic SOC has a different phase for each matrix element. 

If we diagonalize Eq.\ref{eq:4stateHamiltonian}, the relevant energies and wavefunctions are (and a general form for the adiabatic wavefunctions in $n$-dimensions is given in the Appendix):

\begin{equation} 
     E_0 = -A, \quad  \ket{\psi_0} = \begin{pmatrix}  \sin \frac {\theta} 2  \\ -  \frac{1} {\sqrt{3} } \cos \frac{{\theta}}{2}  \\ -  \frac{1} {\sqrt{3}} e^{-i\phi} \cos \frac {\theta} 2  \\-  \frac{1} {\sqrt{3} }  e^{i\phi} \cos \frac {\theta} 2 \end{pmatrix} \qquad  
     E_1 = -A\cos\theta, \quad \ket{\psi_2} = \begin{pmatrix}  0  \\  \sqrt{\frac{2} {3}}   \\ -\frac 1 {\sqrt{6}}   e^{-i\phi} \\-\frac 1 {\sqrt{6}}   e^{i\phi} \end{pmatrix} \qquad  
    \end{equation}

\begin{equation} 
    E_2 = -A\cos\theta,\quad \ket{\psi_1} = \begin{pmatrix}  0  \\ 0 \\ -\frac 1 {\sqrt{2}}  e^{-i\phi}    \\ \frac 1 {\sqrt{2}}  e^{i\phi}   \end{pmatrix} \qquad 
    E_3 = A, \quad \ket{\psi_3} =  \begin{pmatrix} -\cos \frac {\theta} 2  \\ -  \frac{1} {\sqrt{3}} \sin \frac{{\theta}}{2}  \\ -\frac{1} {\sqrt{3}}  e^{-i\phi} \sin \frac {\theta} 2  \\ -\frac{1} {\sqrt{3}}  e^{i\phi} \sin \frac {\theta} 2 \end{pmatrix}   
\end{equation}

For the benefit of the reader, a plot of adiabatic energies  is provided in Fig. \ref{fig:geometry}:

\begin{figure} [H]
    \includegraphics[width=1\columnwidth]{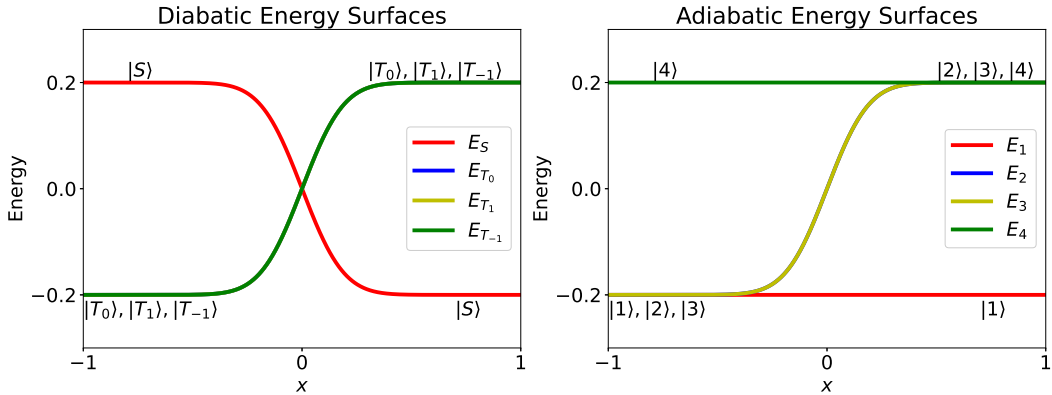}
    \caption{\label{fig:geometry} A schematic figure of the diabatic and adiabatic energy surfaces corresponding to Eq. \ref{eq:4stateHamiltonian}.}  
\end{figure}

\mycomment{Just as for the Hamiltonian in Eq. \ref{eq:2stateHamiltonian}, the diabatic and adiabatic surfaces for the Hamiltonian in Eq. \ref{eq:4stateHamiltonian} are completely flat in the $y$-direction.  In fact,  the upper and lower adiabats here are also flat in the $x$-direction. However,  the middle two adiabats are not flat in the $x$-direction; furthermore, because these states are entirely degenerate at all points in space, there is no unique, well-defined means of distinguishing one from the other.}

Under a change of basis, the Hamiltonian in Eq. \ref{eq:4stateHamiltonian} can be made entirely real-valued:
\begin{equation}\label{eq:4stateHamiltonianreal} 
    \bm {\tilde H} = \bm U^\dagger \bm H \bm U =  A\begin{pmatrix} 
    \cos\theta  &  \frac {\sin\theta} {\sqrt{3}}   &  -\frac {\sqrt{2} \sin\theta \sin \phi } {\sqrt{3}}  &\frac {\sqrt{2} \sin\theta \cos \phi } {\sqrt{3}}   \\
    \frac {\sin\theta} {\sqrt{3}}   & -\cos\theta  & 0 &0 \\ 
    -\frac {\sqrt{2} \sin\theta \sin \phi } {\sqrt{3}} & 0 & -\cos\theta & 0 \\
    \frac {\sqrt{2} \sin\theta \cos \phi } {\sqrt{3}}  & 0 & 0 & -\cos\theta
\end{pmatrix} \end{equation}
where:
\begin{equation}\label{eq:changeofbasis} 
    \bm U =  \begin{pmatrix} 
    1 & 0 & 0 & 0\\
    0 & 1 & 0 & 0 \\
    0 & 0 &\frac {i} {\sqrt 2} & \frac {1} {\sqrt 2} \\
    0 & 0 &-\frac {i} {\sqrt 2} & \frac {1} {\sqrt 2}
\end{pmatrix} \end{equation}

At this point, consider exact wavepacket scatting dynamics for
the degenerate ISC model in Eq. \ref{eq:4stateHamiltonian}.
Although Eq. \ref{eq:bf} might suggest that the Berry force is zero for each state, numerical results in Ref. \citenum{Bian2021} show clearly that a nuclear  wavepacket incident from singlet state $\ket{S}$  will split into three separating daughter wavepackets.  
The daughter wavepacket on $\ket{T_1}$ transmits upward with $p_y = +W$, the daughter wavepacket on $\ket{T_0}$ transmits straight
across with $p_y = 0$, and the daughter wavepacket on $\ket{T_{-1}}$ transmits downward with $p_y = -W$.
The Berry force is clearly nonzero and leads to decoherence for this Hamiltonian, causing a momentum shift when a wavepacket changes its diabat. 

In general, for the Hamiltonian in Eq. \ref{eq:4stateHamiltonian}, we find that the Berry force has the following effect on the final momentum-shift in the  $y$-direction  (depending on the initial and final spin-diabat):

\begin{table}[H] 
    \begin{tabular}{ c  c @{\qquad}c  } 
    \hline
    Initial Diabat& Final Diabat & $p_y$ shift \\
    \hline
    $\ket{S}$, $\ket{T_0}$  &  $\ket{S}$, $\ket{T_0}$ & $0$\\ 
    $\ket{S}$, $\ket{T_0}$ &  $\ket{T_1}$ & $-W$\\ 
    $\ket{S}$, $\ket{T_0}$ &  $\ket{T_{-1}}$ & $+W$\\ 
    $\ket{T_1}$   & $\ket{S}$, $\ket{T_0}$ & $+W$\\ 
    $\ket{T_1}$   & $\ket{T_{-1}}$ & $+2W$\\     
    $\ket{T_{-1}}$   & $\ket{S}$, $\ket{T_0}$ & $-W$\\ 
    $\ket{T_{-1}}$   & $\ket{T_{1}}$ & $-2W$\\ 
    \hline
    \end{tabular}
    \caption{\label{tab:table1} The average momentum shift in the $y$-direction as found in scattering experiments.}
\end{table}

 From a theory perspective, one often refers to the Berry curvature in Eq. \ref{eq:Berrycurvature} as  a non-degenerate Berry curvature; this curvature leads to the adiabatic Berry force in Eq. \ref{eq:bf}. For the present problem, however, where 
the middle two adiabats are degenerate everywhere and three states are degenerate asymptotically, one must  consider the concept of a degenerate Berry curvature \cite{Wilczek1984}:
\begin{equation} \label{eq:nonabelian}
\Omega_{jk}^{\alpha\beta} = i \left(\frac {{\partial  d_{jk}^{\beta} }} {\partial R_\alpha}- \frac { {\partial  d_{jk}^{\alpha}}}  {\partial R_\beta}  \right) - 
i \sum_{l} \left( {d}_{jl}^{\alpha}  { d}_{lk}^{\beta} -   {d}_{jl}^{\beta}  { d}_{lk}^{\alpha}\right)
\end{equation}
 
Whereas Eq. \ref{eq:bf} reflects how a single vector moves in a larger vector space as a function of nuclear coordinates, the expression on the right hand side of Eq. \ref{eq:nonabelian} reflects how a multidimensional subspace evolves within an even large vector space as a function of nuclear coordinate. As such,  Eq. \ref{eq:nonabelian} is meaningful (i.e. nonzero) only when the electronic index $l$ runs over one, two or three states.  If we extend this index to include all four states, we will find that the entire expression is identically zero (which is effectively the curl condition \cite{Mead1982} for a closed electronic space).

As far as we are aware,  there is at present no simple
means of incorporating the concept of a degenerate Berry curvature into a semiclassical, nonadiabatic dynamics algorithm.

\subsection{A "quasi-diabatic" ansatz and an effective "magnetic" field}  \label{sec:quasi-diabat} 

Tully's  FSSH dynamics\cite{Tully1990}  are predicated on the idea that, to best account simultaneously for nuclear barriers, strong and weak electronic coupling \cite{Subotnik2016}, and detailed balance \cite{Schmidt2008},  the optimal approach is to propagate dynamics along adiabatic potential energy surfaces.  That being said,  the phenomenological behavior described in Table \ref{tab:table1} makes clear that there are  limitations to running dynamics along adiabatic surfaces. After all, if one ignores the distinction between $m_s=1$ and $m_s=-1$ triplets, how could one possibly account for the spin-dependent changes of nuclear motion in Table \ref{tab:table1}?  At the same time, as a practical matter, if one works with degenerate states, how does one pick such a state uniquely with a unique derivative coupling?  It would seem that, in order to extend our capacity to simulate nonadiabatic dynamics to degenerate state crossings (like the ISC in Eq. \ref{eq:4stateHamiltonian}),
more information is needed; for instance, it might be natural
to include a diabatic state to help guide each trajectory. But how should such diabats be chosen? And even importantly, how should such a diabatic index actually guide a given nuclear trajectory? We will now address these questions in turn.

First, we consider the question of generating diabats. For an adiabatic electronic state that is largely of singlet character, we define the quasi-diabatic state to be the adiabat. However, for those  adiabatic electronic states that are largely of triplet character, we wish to procure a diabatic representation. So far, for the Hamiltonian in Eq. \ref{eq:4stateHamiltonian}, our most effective idea for generating  diabats is to  rotate the relevant adiabats so as to align  them with the targets triplet diabats using the well-known Kabsch algorithm \cite{Kabsch1976} (which is well-known in many contexts \cite{subotnik:2004:rued, cederbaum:1993:advchemphys}).
 
\begin{figure} [H]
    \centering  \includegraphics[width=0.6\columnwidth]{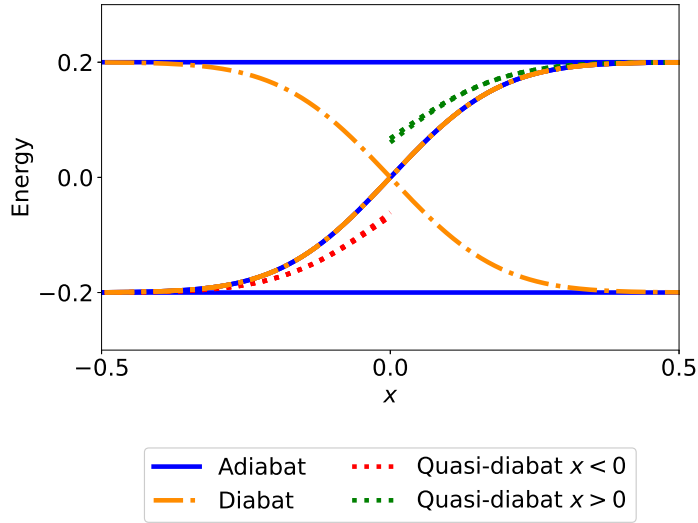}
     \caption{\label{fig:quasidiabatPES} A zoomed-in plot of the energy surfaces in the $x$-direction for the four-state model described by Eq. \ref{eq:4stateHamiltonian}. 
     The solid lines correspond to adiabats, the dash-dotted lines correspond to diabats and the dotted lines correspond to quasi-diabats. 
     The triplet quasi-diabats (red and green dotted) are always nearly triply degenerate and will coincide with the diabats asymptotically. 
     The singlet quasi-diabat is the same as original singlet diabat.}
 \end{figure}

The resulting electronic states are clearly {\em not} entirely diabatic, and so we call this new basis a "quasi-diabatic" basis. The basis is plotted in Fig. \ref{fig:quasidiabatPES}.
 
Mathematically, the process described above works as follows. 
Asymptotically, as $x \to \infty$ case, the adiabats $\ket{2}$,$\ket{3}$ 
and $\ket{4}$ coincide with diabats $\ket{T_0}$,$\ket{T_1}$ and $\ket{T_{-1}}$. Let ${\bm R}$ be the $3
\times 3$ rotation matrix that rotates  $\{\ket{2} , \ket{3}, \ket{4}\}$ into 
the desired quasi-diabats, $\{ \ket{ {\tilde T}_0}, \ket{{\tilde T}_1},\ket{{\tilde T}_{-1}} \}$:
\begin{equation} \{ \ket{ {\tilde T}_0},\ket{{\tilde T}_1}, \ket{{\tilde T}_{-1}} \} = \{ \ket{2}, \ket{3}, \ket{4} \} {\bm R}^\dagger  \end{equation}
To best  align the adiabats with the diabats in this limit, the rotation matrix $\bm R$ is defined to be the matrix that maximizes
the overlap:
\begin{eqnarray}
f(R) = \left<{\tilde T}_0 |T_0\right> +
\left<{\tilde T}_1 |T_1\right> +
\left<{\tilde T}_{-1} |T_{-1}\right> 
\end{eqnarray}
Let $\bm P$ by the $4 
\times 3$ matrix of column vectors that span the relevant adiabatic space, $\bm P = 
\left[\ket{2}, \ket{3}, \ket{4} \right]$, and let ${\bm Q}$ be the matrix of column vectors spanning the relevant diabatic space,
${\bm Q} = \left[\ket{T_0}, \ket{T_1}, \ket{T_{-1}}\right]$
The function to be optimized is:  
$  f({\bm R}) = {\rm {Tr}}  (( {\bm R}{\bm P}^\dagger )  {\bm Q})$. 
It has been proven repeatedly in the literature \cite{Kabsch1976} that $f({\bm R})$ is maximal when: 

\begin{equation} R = {\bm O}^\dagger ({\bm O} {\bm O}^\dagger)^{-\frac 1 2} \end{equation}
where $\bm O = {\bm P}^\dagger   {\bm Q}$.
For the given Hamiltonian in Eq. \ref{eq:4stateHamiltonian}, the quasi-diabats in the region $x>0$ can be calculated analytically:
\begin{equation} 
    \ket{ {\tilde T}_0} = \begin{pmatrix}  \frac 1 {\sqrt 3} \cos \frac {\theta } 2   \\  \frac 2 3 + \frac 1 3 \sin  \frac {\theta } 2      \\  - \frac {e^{-i\phi}} {3} (1 - \sin  \frac {\theta } 2  ) \\  - \frac {e^{i\phi}} {3} (1 - \sin  \frac {\theta } 2  ) \end{pmatrix},  
    \ket{ {\tilde T}_1} = \begin{pmatrix}  \frac {e^{i\phi}}  {\sqrt 3}    \cos \frac {\theta } 2 \\ - \frac {e^{i\phi}} {3} (1 - \sin  \frac {\theta } 2  )   \\ \frac 2 3 + \frac 1 3 \sin \frac {\theta} 2 \\  -\frac {e^{2i\phi}} 3 (1 - \sin \frac {\theta }2 ) \end{pmatrix},  
    \ket{ {\tilde T}_{-1}} = \begin{pmatrix}  \frac {e^{-i\phi}} {\sqrt 3}    \cos \frac {\theta } 2 \\ - \frac {e^{-i\phi}} {3} (1 - \sin  \frac {\theta } 2  )  \\  -\frac {e^{-2i\phi}} 3 (1 - \sin \frac {\theta }2 ) \\ \frac 2 3 + \frac 1 3 \sin \frac {\theta} 2   \end{pmatrix}, 
\end{equation}
Similarly, the quasi-diabats in the region $x<0$ can be calculated analytically:
\begin{equation} 
    \ket{ {\tilde T}_0} = \begin{pmatrix}  -\frac 1 {\sqrt 3} \sin \frac {\theta } 2   \\  \frac 2 3 + \frac 1 3 \cos  \frac {\theta } 2      \\  - \frac {e^{-i\phi}} {3} (1 - \cos  \frac {\theta } 2  ) \\  - \frac {e^{i\phi}} {3} (1 - \cos  \frac {\theta } 2  ) \end{pmatrix},  
    \ket{ {\tilde T}_1} = \begin{pmatrix}  \frac {e^{i\phi}}  {\sqrt 3}    \sin \frac {\theta } 2 \\ - \frac {e^{i\phi}} {3} (1 - \cos  \frac {\theta } 2  )   \\ \frac 2 3 + \frac 1 3 \cos \frac {\theta} 2 \\  -\frac {e^{2i\phi}} 3 (1 - \cos \frac {\theta }2 ) \end{pmatrix},  
    \ket{ {\tilde T}_{-1}} = \begin{pmatrix}  \frac {e^{-i\phi}} {\sqrt 3}    \sin \frac {\theta } 2 \\ - \frac {e^{-i\phi}} {3} (1 - \cos  \frac {\theta } 2  )  \\  -\frac {e^{-2i\phi}} 3 (1 - \cos \frac {\theta }2 ) \\ \frac 2 3 + \frac 1 3 \cos \frac {\theta} 2   \end{pmatrix}, 
\end{equation}
Note that there is a discontinuity in the quasi-diabats at $x=0$ using this ansatz.

Next, let us consider the question of how to best employ these quasi-diabats.  To that end, we will make the ansatz that each trajectory will carry both an adiabatic index and a quasi-diabatic index, and then one can simply apply the Berry force for the associated quasi-diabatic  to the given nuclear trajectory. In other words, our approach will be to calculate how each quasi-diabat changes with nuclear geometry using the non-degenerate "Berry curvature" tensor for each quasi-diabat following Eq. \ref{eq:Berrycurvature} (rather than the more complicated degenerate Berry curvature tensor in Eq. \ref{eq:nonabelian}), and then applying the corresponding magnetic field. A calculation yields (for $x>0$):
\begin{equation} {\Omega}^{xy}_{{\tilde T}_0} = i (\nabla_x \bra{ {\tilde T}_0} \nabla_y \ket{ {\tilde T}_0} - \nabla_y \bra{ {\tilde T}_0} \nabla_x  \ket{ {\tilde T}_0}) = 0 \end{equation} 
\begin{equation} {\Omega}^{xy}_{{\tilde T}_1}  = \frac 1 3 \nabla_x \theta \nabla_y \phi \cos \frac {\theta} 2   \end{equation} 
\begin{equation} {\Omega}^{xy}_{{\tilde T}_{-1}}  =  -\frac 1 3 \nabla_x \theta \nabla_y \phi \cos \frac {\theta} 2  \end{equation} 
Note that, for any and all of the three quasi-diabats of interest,  ${\bm \Omega}^{xy} = -{\bm \Omega}^{yx}$ and ${\bm \Omega}^{xx} = {\bm \Omega}^{yy} = 0$.  
Next, we address the singlet quasi-diabat  $\ket{\tilde{S}}$.  In the regime where $x \to \infty$,
\begin{equation} 
    \ket{\tilde S} = \begin{pmatrix}  -\cos \frac {\theta } 2   \\   -\frac 1 {\sqrt 3} \sin  \frac {\theta } 2      \\  - \frac {e^{-i\phi}} {\sqrt 3}  \sin  \frac {\theta } 2  \\  - \frac {e^{i\phi}} {\sqrt 3}  \sin  \frac {\theta } 2   \end{pmatrix}, 
\end{equation}
and  the Berry curvature is always zero, $\stackrel{\leftrightarrow}{ \Omega}_{S}$ = 0.
At this point, we can define a Berry force of interest (on either side of $x =0$):
\begin{equation}\label{eq:degeneratebf}  \vec {F}^{Berry}_{\mu} =  \eta \vec {\Omega}_{\mu} \cdot \frac {\vec {p}} m =  \eta  (i\vec { \nabla} \times \bra{\mu} \vec {\nabla} \ket{\mu}) \cdot \frac {\vec {p}} m \end{equation}
or in index form:
\begin{eqnarray}\label{eq:degeneratebfindex}  
F^{Berry, \alpha}_{\mu} =  i\eta \sum_{\beta}\left(  \nabla_{\alpha}   \bra{\mu} \nabla_{\beta} \ket{\mu} -  \nabla_{\beta}   \bra{\mu} \nabla_{\alpha} \ket{\mu} \right) \frac {p_{\beta}} m  
\end{eqnarray} 
Here, for reasons to be explained below, we have introduced a constant $\eta$. $\mu$ is the active quasi-diabat index. 

Eq. \ref{eq:degeneratebf} above has many features of interest as far as replicating the desired changes in scattering momenta according to Table \ref{tab:table1}. For instance, note that,  if one introduces a universal {\em ad hoc} scaling factor $\eta = 3/2$ in Eq. \ref{eq:degeneratebf} and one evaluates a one-dimensional integral analogous to Eq. \ref{eq:integral}, one can calculate the the momentum shift in the $y$-direction for a trajectory along quasi-diabat $\tilde{T}_1$ as follows:
\begin{equation} \begin{aligned}  \label{eq:integralq}
\Delta p^y &= \int_{t=0}^{t = \infty} \frac {3} {2m} {\Omega}_{T_1}^{yx} p^x dt = \int_{0}^{\pi} \frac {1} {2m}  \nabla_y\phi \cos\frac {\theta} 2 d\theta = W
\end{aligned}\end{equation}
In other words, a momentum shift  $W$ appears. Note that the integral in Eq.  \ref{eq:integralq} is not relevant for any surface hopping calculation because, as mentioned above (and see Fig. \ref{fig:quasidiabatPES}), the definition of each quasi-diabat changes discontinuously at $x=0$ (whereas in Eq. \ref{eq:integralq} we have simply integrated the Berry force for $x<0$ over the entire $x$-axis).  Nevertheless, if we imagine a trajectory that is initialized on the upper 
singlet $\ket{S}$ at $\theta = 0$ and then transmits to $\ket{T_1}$ at $\theta = \pi$,  if we treat the discontinuity at $x=0$ properly, it should be clear that we can integrate the Berry forces in Eq. \ref{eq:degeneratebf} to 
yield the proper outgoing momentum. 
As a side note, note that if one enters on diabat $\ket{T_1}$ on the left and one exits on diabat $\ket{T_1}$ on the right, there should be no momentum shift. Thus, again, the integral in Eq. is not directly relevant.  Encouragingly, we find that this physical result can be captured if we integrate the Berry force for $x<0$ over the region $x \in [-\infty,0]$ and the Berry force for $x>0$ over the region  $x \in [0, \infty]$; the two Berry forces cancel exactly as they should.

\subsection{A modified fewest switches surface hopping algorithm} 
 At this point, we can introduce our proposed surface hopping protocol.
To begin our discussion, and to introduce the necessary notation, let us briefly review  our FSSH algorithm for a non-degenerate system with 
a Berry force.  As usual \cite{Tully1990}, we imagine initialization of a swarm of independent trajectories 
based on a Wigner distribution. For each trajectory, the nuclear motion is propagated classically along a 
single active adiabatic surface $j$ with equations of motion:
\begin{equation} \dot {\vec {r}} = \frac {\vec { p}} m \end{equation}
\begin{equation} \dot {\vec {p}} = -\vec {\nabla} E_j + \vec { F}^{Berry}_j \end{equation}

The adiabatic electronic amplitude ${\bm c}$ is propagated by the Schr\"{o}dinger equation:
\begin{equation} \dot  c_k = - \frac i {\hbar} E_k c_k - \sum_j {\frac {\vec {p}} m} \cdot {\vec { d}}_{jk} c_k \end{equation}
According to Tully's original paper\cite{Tully1990}, a trajectory on active surface $j$ randomly hops to surface $k$ with rate:
\begin{equation}  \label{eq:hoprate}  g_{j\to k} = \max \left[2 {\rm Re} \left({\frac {\vec {  p}} m} \cdot {\vec { d}}_{jk} \frac {\rho_{kj}} {\rho_{jj}} \right) \Delta t , 0  \right] \end{equation}
\mycomment{This hopping rate was originally guessed by Tully so as to enforce consistency between the fraction of trajectories on surface $j$ and the population $|c_j|^2$; the validity of this ansatz was later demonstrated by comparison with the quantum classical Liouville equation (QCLE)\cite{Subotnik2013, Kapral1999}.}
When a hop occurs, one rescales the momentum to conserve energy in \mycomment{the direction of the derivative coupling (assuming it is real). }
%(see Sec \ref{sssec:rescale}).
  
\subsubsection{Hops Between Quasi-diabats} \label{sssec:hopqd}
In order to extend the FSSH algorithm to a  system with degenerate electronic states, we have argued that, in addition to an active adiabat (which yields Born-Oppenheimer force information), each trajectory should also carry an active quasi-diabat (which yields  Berry force information).   
\mycomment{However, it is not possible to simply assign a quasi-diabatic index to a trajectory in the same way that we assign an adiabatic index. After all, such an active quasi-diabat index would face two incompatible requirements: one would require the active quasi-diabat to be consistent with both the electronic wavefunction as well as the active adiabatic (which is impossible!). For this reason, our approach instead will be to assign a quasi-diabat only when the active adiabat is mostly of triplet character; in such a case, one can use the quasi-diabat as a means of quantifying which $m_s$ level is being  populated by a given trajectory. Thereafter, if one wishes to use a quasi-diabatic index to sample the correct distribution of $m_s$ triplets so that a swarm of trajectories moves with a meaningful sampling of Berry forces (consistent with the electronic amplitudes), one must also allow for the quasi-diabat index of a given trajectory to hop between electronic states as a function of time. Obviously,  this hopping rate must be chosen very carefully so as to maintain all of the consistency checks above.}

With these caveats in mind, it is straightforward to imagine three scenarios whereby a given trajectory will need to hop between active quasi-diabats:
\begin{enumerate} 
\item  Whenever an adiabatic hop occurs, the active quasi-diabat may change, and one must carefully assess the situation.
For instance, if a trajectory hops from adiabat $\ket{1}$ to adiabat $\ket{4}$ in the region $x < 0$, 
the active quasi-diabat must also necessarily change to $\ket{S}$.   In general and more precisely, if the new adiabat is primarily of singlet character (or formally maps to a singlet quasi-diabat), then one must also switch the active quasi-diabat to the singlet. If the new adiabat is primarily of triplet character, then one must switch the active quasi-diabat to one of three triplets (and the relative ratio of each triplet is calculated according to the electronic amplitude). 

\item The active quasi-diabat must also change due to 
the evolution of the electronic wavefunction in order to keep consistency between the electronic amplitudes and the choice of active surface.  Mathematically, we calculate the stochastic  quasi-diabatic hopping probability according to the fewest switches hopping rate. Thus, the hopping probability from active quasi-diabat $a$ to  another quasi-diabat $b$ is calculated in the quasi-diabatic basis by the expression:
\begin{equation} {\label{eq:qdhoprate}}g_{a\to b} = \max \left[\frac {T_{ba}} {\rho_{aa}}   \Delta t , 0  \right]. \end{equation}
where:
\begin{equation} T_{ba} = 2 {\rm Im} \left( V_{ab} \rho_{ba}\right) - 2 {\rm Re} \left({\frac {\vec { p}} m} \cdot {\vec { d}}_{ba} {\rho_{ab}}   \right)\end{equation}

\item  When a trajectory transmits through the crossing point 
(at $x = 0$ for our model of Eq. \ref{eq:4stateHamiltonian}) on adiabat $\ket{1}$ or $\ket{4}$, the mapping between adiabats and quasi-diabats changes because of our discontinuous choice of quasi-diabats. 
For example, for a  trajectory transmitting on adiabat $\ket{4}$ from $x = -\infty$ to $x = \infty$, the active quasi-diabat must change from $\ket{S}$ ($x<0$) to one of three quasi-diabats after crossing ($x>0$). Thus, whenever a trajectory passes $x = 0$ on $\ket{1}$ or $\ket{4}$, we must check for a quasi-diabatic hop. As above, if the new adiabat is primarily of singlet character, the new quasi-diabat is the corresponding singlet quasi-diabat.  If the new quasi-diabat is of triplet character, then the new quasi-diabat is set to one of the triplet quasi-diabats (and the exact choice is again chosen stochastically according  to the electronic amplitude).
\end{enumerate}

\begin{figure} [H]
    \centering
    \includegraphics[width=0.6\columnwidth]{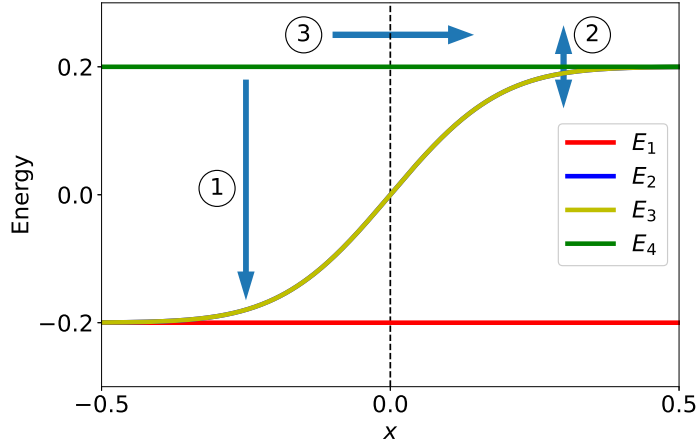}
    \caption{\label{fig:hops} Schematic diagram for three scenarios of quasi-diabatic hops: (1) an adiabatic hop leading to a quasi-diabatic hop, (2) a hop between triplet quasi-diabatic states, and (3) a trajectory transmitting through the crossing point (at x = 0).} 
\end{figure}

\subsubsection{Hops Between Adiabats}
An adiabatic hop may lead to a quasi-diabat hop. That being said, if the adiabatic hop is between a nearly degenerate group of states (i.e. both the initial and final states have triplet character), then no quasi-diabat hop is needed.  In such a case, we propose that all the rules for hopping between adiabats as prescribed by Tully\cite{Tully1990} are unchanged. For example, the hopping probability is calculated by Eq. \ref{eq:hoprate}.
\subsubsection{\label{sssec:rescale} Momentum Rescaling}

We now return to the thorny question of momentum rescaling. \mycomment{Note that, for a complex-valued Hamiltonian, with complex-valued derivative coupling, there is no unique direction for rescaling momentum. While the derivative coupling is clearly the correct momentum rescaling direction for standard, non-degenerate  real-valued Hamiltonians \cite{Herman1984, Subotnik2013, Kapral1999, Kapral2016}, the notion of a complex-valued momentum is not practical or even well defined (because the derivative coupling is only defined up to a phase [or a gauge]).  Moreover, this same problem (i.e. the lack of an obvious, unique momentum rescaling direction) arises whenever
one encounters any nonadiabatic problem with electronic degeneracy, e.g. a singlet-triplet Hamiltonian.\cite{Bian2021}
For these reasons, in recent years, we have spent a great deal of time analyzing the question of momentum rescaling.}

According to Ref. \citenum{Wu2021}, the optimal rescaling direction for the model Hamiltonian described in Eq. \ref{eq:2stateHamiltonian} is a combination of  two unique directions ($\hat h$ and $\hat k$) which are functions of $\nabla \theta$ and $\nabla \phi$; for the present Hamiltonian, these are merely the $x$- and $y$-directions. The basic idea in Ref.  \citenum{Wu2021} is that, whenever an adiabatic hop occurs, one adjusts the momentum in order to ensure that the trajectory leaves the crossing region with the correct asymptotic momentum as induced by the correct Berry force. Given the encouraging results found in Ref. \citenum{Wu2021} for the two state problem, we are hopeful that a similar idea will apply for the present  (degenerate) model. \mycomment{Note that, according to Refs. \cite{Miao2019, Wu2021}, if one does not use an optimal momentum rescaling direction, FSSH simply does not predict accurate answers for the Hamiltonian in Eq. \ref{eq:2stateHamiltonian}}

At this point, let us be more precise. According to Ref. \citenum{Wu2021}, the momentum rescaling procedure contains two steps: First, one solves for the target momentum in the $y$-direction in order to fix the asymptotic momentum shift. Second, one  then solves for the target momentum in the $x$ direction according to energy conservation. Now, in our current singlet-triplet model, the Berry force follows the active quasi-diabatic index (rather than the adiabatic index), and so one must presume that momentum rescaling must be applied both for adiabatic and for quasi-diabatic hops. With regards to adiabatic hops, there is a very easy protocol to follow. If an adiabatic hop occurs without a quasi-diabat hop, we simply rescale momentum in the $x$-direction (corresponding to the $\nabla \theta$ direction in Eq. \ref{eq:4stateHamiltonian}); this protocol make sense because
the Berry force is tied to the quasi-diabatic index (not the adiabatic index). The more interesting question is how to rescale velocities when there is a hop  between quasi-diabats.

To that end, consider a hop between quasi-diabats. We focus on the momentum correction $\Delta p^y$.   After a switch from quasi-diabat $a$ to quasi-diabat $b$ at time $t$, the momentum correction $\Delta p^y$ must satisfy the following relation to guarantee the correct final momentum-shift,
\begin{equation}\label{eq:rescale} p^y [a | 0 \to t]   + p^y [b |t \to \infty]  +\Delta p^y  =   p^y [a \to b | 0 \to \infty] \end{equation}
 
\noindent where $p^y [a\to b | 0 \to \infty]$ denotes the asymptotic momentum change (which can be found in Table \ref{tab:table1}) for a trajectory incident from quasi-diabat $a$ and leaving the crossing region on quasi-diabat $b$. $p^y [a| 0 \to t]$ denotes the momentum induced by Berry force  from time $0$ to $t$ for a trajectory on quasi-diabat $a$ . Notice that Eq. \ref{eq:rescale} is a bit complicated  because of the discontinuity of the quasi-diabats around $x = 0$ ($\theta = \frac \pi 2$). A calculation reveals the following two scenarios (where we define $t''$ as that time where the trajectory crosses the line $\theta=\pi/2$):

\noindent Case 1: $ \theta(t) <\frac{\pi}{2}$
\begin{equation}
\begin{aligned}
    &p^y [a | 0 \to t] = \int_{0}^{t} F_{a,(\theta<\pi/2) }^{Berry,y}  dt' \\
    &p^y [b | t \to \infty] = \int_{t}^{t''} F_{b,(\theta<\pi/2) }^{Berry,y}  dt' + \int_{t''}^{\infty}  F_{b,(\theta>\pi/2) }^{Berry,y} dt'  
 \end{aligned}
\end{equation}
 
\noindent Case 2: $ \theta(t) > \frac{\pi}{2}$
\begin{equation}
\begin{aligned}
&p^y [a | 0 \to t] = \int_{0}^{t''} F_{a,(\theta<\pi/2) }^{Berry,y}  dt + \int_{t''}^{t}  F_{a ,(\theta>\pi/2) }^{Berry,y} dt' \\
&p^y [b | t \to \infty] =  \int_{t'}^{\infty}  F_{b,(\theta>\pi/2) }^{Berry,y} dt '
 \end{aligned}
\end{equation}

In terms of the singlet-triplet Hamiltonian in Eq. \ref{eq:4stateHamiltonian}, all of the  integrals above are merely functions of $\theta$ so that the momentum shift $\Delta p_y$ can be locally determined and in fact is a piece-wise continuous function of $\theta$ (as shown in Table 2). 
Note that momentum corrections in Table 2 apply to all possible quasi-diabatic hopping scenarios as discussed in Sec. \ref{sssec:hopqd}:

\begin{table}[H] 
    \centering
    \begin{tabular}{ c  @{\qquad}c @{\qquad}c  } 
    \hline
    $\ket{a} \to \ket{b}$ & $\Delta p_y (\theta < \frac {\pi} 2)$ & $\Delta p_y  (\theta > \frac {\pi} 2)$ \\
    \hline  
    $\ket{\tilde S}$, $\ket{\tilde{T}_0} \to \ket{\tilde{T}_0}, \ket{\tilde{S}}$& 0 & 0\\ 
    $\ket{\tilde S},  \ket{\tilde{T}_0} \to  \ket{\tilde{T}_1}$ &  $-\sin \frac {\theta} 2 W$  &  $-\cos \frac {\theta} 2 W$\\ 
    $\ket{\tilde S},  \ket{\tilde{T}_0} \to  \ket{\tilde{T}_{-1}}$ &  $\sin \frac {\theta} 2 W$  &  $\cos \frac {\theta} 2 W$\\ 
    $\ket{\tilde{T}_1} \to \ket{\tilde S}, \ket{\tilde{T}_0}$ & $ \cos \frac {\theta} 2W$  & $ \sin \frac {\theta} 2 W$\\ 
    $\ket{\tilde{T}_{-1}} \to \ket{\tilde S}, \ket{\tilde{T}_0}$ & $-\cos \frac {\theta} 2W$  & $ -\sin \frac {\theta} 2 W$\\   
    $\ket{\tilde{T}_{1}} \to \ket{\tilde{T}_{-1}}$   & $2\cos \frac {\theta} 2W$ &  $-2\sin \frac {\theta} 2W$\\ 
     $\ket{\tilde{T}_{-1}} \to \ket{\tilde{T}_{1}}$  & $-2\cos \frac {\theta} 2W$ &  $2\sin \frac {\theta} 2W$\\ 
    \hline
    \end{tabular}
    \caption{\label{tab:table2} The momentum corrections in the  $y$-direction that must be applied whenever there is a quasi-diabatic hop.}
\end{table}

\subsection{Momentum Reversal}

At this point, one more nuance must be addressed. Often, during a hop, one finds that the hop is frustrated because there is not enough energy to achieve rescaling in the chosen direction. Indeed, there is a large literature on the subject of momentum reversal upon a frustrated hop  \cite{Jasper2001,Jasper2003} in the context of real-valued Hamiltonians. In general, the most useful momentum reversal criteria suggested by Jasper and Truhlar \cite{Jasper2003} is:
\begin{equation} ({\vec { F}_{\lambda}} \cdot \vec { d}_{\lambda j})  (\vec { p} \cdot \vec { d}_{\lambda j}) < 0 
\end{equation}

\mycomment{Velocity reversal was found to be pivotal in Ref. \citenum{Wu2021}, and the} issue becomes even more essential for the present singlet-triplet Hamiltonian. When a quasi-diabatic hop occurs,  there can be a large probability that the  momentum rescaling procedure will be frustrated (because the presence of a Berry force can demand a large change of momentum). 
Empirically, based on comparison with exact solutions, we have found several rules as to when we ought to reverse momenta. On the one hand,  if a trajectory is incident from a lower triplet or an upper singlet, optimal results require momentum reversal upon a frustrated quasi-diabatic hop. On the other hand,  for trajectories that are incident from a lower singlet or an upper triplet, we recover the best results if we do not not reverse the momentum following a quasi-diabatic hop. In other words, for the flat model Hamiltonian in Eq.  \ref{eq:4stateHamiltonian}, we implement momentum reversal after a frustrated quasi-diabatic hop if the trajectory's momentum is in  the same direction as the adiabatic force for the middle pair of degenerate adiabats: 
\begin{equation} \label{eq:reverse} 
\vec {p} \cdot \vec { F}_{\rm middle} < 0 \end{equation}
For a frustrated adiabatic hop, the momentum reversal is applied according to Ref. \citenum{Wu2021} when: 
\begin{equation} (\vec { F}_\lambda \cdot \vec \nabla \theta) (\vec  { p} \cdot \vec \nabla \theta) < 0 \end{equation} 

\subsection{The Extreme Diabatic Limit}
The case of an extreme diabatic crossing represents a problem for any surface hopping algorithm with Berry force: after all, even though one wishes to apply a  Berry force to ensure the correct asymptotic momentum for the rare trajectory that changes diabat, one is also aware that most trajectories do not switch diabats -- and so there is no reason to implement a Berry force which can only lead to worse results. In particular, in the extreme diabatic limit,  our experience with a two state crossing shows that including a  Berry force can sometimes lead to a large  overestimation of
reflection, as trajectories (that should hop to another adiabat [i.e. stay on the same diabat]) are reflected too early because of Berry force. In the present paper, we find similar results for a  singlet-triplet crossing model. Thus, \mycomment{whenever a trajectory reaches the crossing point (at $\theta = \frac \pi 2$ and $x = 0$)}, our protocol below is to turn off the Berry force and ignore the momentum rescaling in Eq. \ref{eq:rescale} if the hop is in the extreme diabatic limit: 

\begin{equation} \label{eq:extreme}  \sum_{\lambda \neq j}  \abs{ \frac {  \left(\vec d_{\lambda j} \cdot  \frac {\vec p} m \right)}  {\Delta E_{\lambda j}} } > 2 \end{equation}
This procedure makes sense because, in the extreme diabatic limit, we can be certain that there will be a hop ahead over the next few time steps.
A more general method for  treating the extreme diabatic limit in surface hopping will be discussed in a future publication.

\subsubsection{Outline of a degenerate FSSH algorithm}
At this point, we can formally summarize our proposed singlet-triplet nonadiabatic FSSH algorithm. In general, we use all of the numerical tricks from Refs. \citenum{Jain2016, Wu2021}.
\begin{enumerate}[Step 1:]
    \item We initialize the classical position $\vec r$, momentum $\vec p$, electronic amplitude  $\bm c$ and active adiabat $\lambda$  
    for an ensemble of trajectories according to a given initial condition (and usually involving a Wigner transform). 
    \item We assign an active quasi-diabat  $\mu$  to each trajectory. If $\lambda$ is of singlet character, we  
    assign $\mu$ to be a singlet. Otherwise, if $\lambda$ is of triplet character, we generate a quasi-diabatic basis and assign $\mu$ to be one of the triplet quasi-diabats; the assignment of which triplet is made stochastically
    according to the electronic population in a quasi-diabatic basis. 

    \item \label{steprp} We propagate $\vec  r$, $\vec p$  from current time $t = t_0$ using a classical time step $dt_c$ to $t = t_0 + dt_c$ following the equation of motion
    \begin{equation}  \dot {{\vec r}} =  \frac {\vec { p}} {m} \end{equation}
    \begin{equation}  \label{pdot} \dot { {\vec p}} = -\vec \nabla E_j + \vec { F}^{Berry}_{\mu}  \end{equation}
    
    \item  We calculate a quantum time step $dt_q$ that is small enough to treat sharp trivial crossing accurately and we loop over quantum time steps\cite{footnote4}.

    \begin{enumerate}
        \item We propagate the electronic amplitude $\bm c$ in the adiabatic basis for a quantum time step $dt_q$ according to the Schrodinger equation.   
    
        \item We calculate the hopping probability $g_{\lambda \rightarrow j}$ from active adiabat $\lambda$ to all other adiabatic states $j$ during each quantum time step $dt_q$ using the standard formula in Eq. \ref{eq:hoprate}.  Generate a random number $\zeta$ and if $\zeta > g_{\lambda \rightarrow j}$, attempt a hop. 
        
        (i) If both $j$ and $\lambda$ are mainly triplets, switch $\lambda$ to $j$ (and do not adjust the quasi-diabat index $\mu$). 
        
        (ii) If $j$ is mainly the singlet and $\lambda$ is a triplet, switch $\lambda$ to $j$ and switch $\mu$ to the  singlet.
        
        (iii) If $j$ is mainly a triplet and $\lambda$ is mainly the singlet, switch $\lambda$ to $j$ and switch $\mu$ randomly to one of the triplet quasi-diabats according to the  electronic amplitude.
        
        For any of these three scenarios, rescale the momentum-- first in the $y$-direction following Eq. \ref{eq:rescale}, and then in the $x$-direction according to energy conservation. If the hop is frustrated,  no switch of the adiabats or diabats is invoked.  \cite{footnote2} Move on to step (\ref{c}).
    
        \item \label{c} If the active quasi-diabat $\mu$ is a triplet, we calculate the probability for a purely quasi-diabatic hop from quasi-diabatic triplet $\mu$ to quasi-diabatic triplet $a$ according to Eq. \ref{eq:qdhoprate}. Again, we 
        generate a random number $\zeta$ and attempt a hop if $\zeta > g_{\mu a}$. 
        If a hop is prescribed, as above, we rescale the momentum first in the $\nabla \phi$ ($y$) direction following Eq. \ref{eq:rescale}, then in the  $\nabla \theta$ ($x$) direction to conserve energy. If the hop is frustrated,  the diabatic hop is rejected.
     
        \item Change the time to $t = t+dt_q$. If $t \ge t_0+ dt_c$  continue; else return to step 4(a) and iterate.
    \end{enumerate}

    \item  Check if the trajectory passes through the crossing point (for this model, $x = 0$) within the classical time step.  If so, check if a quasi-diabatic change is needed according to the rules in Sec. \ref{sssec:hopqd}. If not, return to Step \ref{steprp}.
    
    \item At the crossing point, check if the trajectory is in the extreme diabatic limit, i.e. if     $\sum_{\lambda \neq j}  \abs{  \left( \vec d_{\lambda j} \cdot \vec p / m \right) / {\Delta E_{\lambda j}}} > 2 $. 
    \begin{enumerate}
    \item If the hop is in the extreme diabatic limit,  turn off the Berry force in Eq. \ref{pdot} for the remainder of the trajectory. Return to step \ref{steprp} and iterate. 
    \item If the hop is not in the extreme diabatic limit, rescale the momentum following Eq. \ref{eq:rescale}.  
    \end{enumerate}
    If the hop appears to be frustrated, check if  $\vec { p} \cdot \vec { F}_{\rm middle} > 0$. 
    If this condition is true, reverse  the momentum in the $\nabla\theta$ ($x$) direction. 
 
    % \item  Check if the trajectory is  in the extremely diabatic limit, i.e. if  that XXX
    % $\sum_{\lambda \neq j}  \abs{  \left( \vec d_{\lambda j} \cdot \vec p / m \right) / {\Delta E_{\lambda j}}} > 2 $. If so, for later iterations, turn off the Berry force in step 3 and  rescale the momentum in the $\nabla\theta$ ($x$) direction for all hops in step 4 and step 5. 
    % Return to step 3 and iterate.
    
\end{enumerate}
 
In practice, there is one more item that the reader/user must be aware of.  Notice that, by definition, the triplet quasi-diabats are not continuous near $x = 0$, i.e. the composition of quasi-diabats change from $\{\ket{1}, \ket{2} , \ket{3}\}$  to  $\{\ket{2}, \ket{3} , \ket{4}\}$ (or vice versa). See Fig. \ref{fig:geometry}. As a result of this discontinuity, the components of the electronic amplitudes in the quasi-diabats change dramatically at $x=0$.  To treat this discontinuity in a stable fashion, we treat this point as  a trivial crossing problem and simply follow the protocol in Ref. \citenum{Jain2016}. In particular, we optimize the phases of the quasi-diabats across the $x=0$ divider and, by numerically calculating the derivative coupling as the matrix log of the overlap \cite{HammesSchiffer1998}, we generate a finite hopping probability that can be easily integrated over the smaller quantum time steps that compose the larger classical time step.   Note that, in general, for the current problem with degenerate electronic states, one must be careful when calculating the derivative coupling -- if one does not line up phase correctly or if one attempts to use Hellman-Feynman theory for the derivative coupling, one is doomed to either instability or failure (or both).

\section{\label{sec:Results}Results}
The algorithm above has been implemented and run for the Hamiltonian in Eq. \ref{eq:4stateHamiltonian}.  Below, we will compare the results from (i) standard FSSH, (ii) our modified FSSH algorithm, and (iii) exact quantum dynamics. 

\mycomment{For this Hamiltonian, in a rough sense, the Massey parameter is effectively $\frac{2\pi A}{\hbar B v_x}$ where $v_x$ is the velocity in the $x-$direction. For this reason, we will present results for both small and large $A$ values, so that we can investigate both nonadiabatic and adiabatic dynamics. In the extreme adiabatic limit (where the character of the electronic state changes), we expect Berry force effects to be very important (and Tully hopping to be less important).  In the extreme nonadiabatic limit (where the character of the electronic state does not change), we expect Berry force to be less important (and Tully hopping to be more important). In between these two limits, we expect that both dynamical effects (hopping and Berry forces) will be important.   Moreover, as far as the parameter $W$ is concerned, the most interesting physics should arise when $W$ is roughly the same order of magnitude as $v_x$. For this reason, we will show scattering results for a host of different velocities.}

For exact quantum dynamics, the wavepacket 
is propagated using the fast Fourier transform method  \cite{Kosloff1983}. The initial Gaussian wavepacket begins on 
a pure diabatic state:   
\begin{equation} \label{eq:wp} \ket{\Psi({\vec r})} =  \exp \left(-\frac {({\vec r} - {\vec r_0})^2} {\sigma^2} +  \frac {i{\vec p}_0 \cdot {\vec r}} {\hbar}  \right) \ket{\psi_i} \end{equation}
Here $\sigma$ is the width,  ${\vec p}_0 = (p^x_{init}, p^y_{init})$ is the initial momentum, ${\vec r}_0 = (-4, 0)$ is the 
initial position of the wavepacket and $\ket{\psi_i}$ is the initial diabatic electronic state index. The nuclear mass is set to be 1000 and all quantities are in atomic units.   
 
For standard and modified surface hopping results, $2 \times 10^3$ trajectories are sampled from the Wigner probability arising from Eq. \ref{eq:wp}.
(i.e. both $\vec r$ and $\vec p$ are sampled from a Gaussian distribution with $\sigma_x = \sigma_y = 1$ and $\sigma_{p_x} = \sigma_{p_y} = 1/2$ ). 
When propagating the electronic amplitudes, we use the analytical eigenvectors of the Hamiltonian. For the case of initialization on a triplet, the active initial adiabat is randomly chosen according to the initial electronic amplitude.

\subsection{Initialize on the singlet state}
\begin{figure} [H]
    \centering
    \includegraphics[width=0.75\columnwidth]{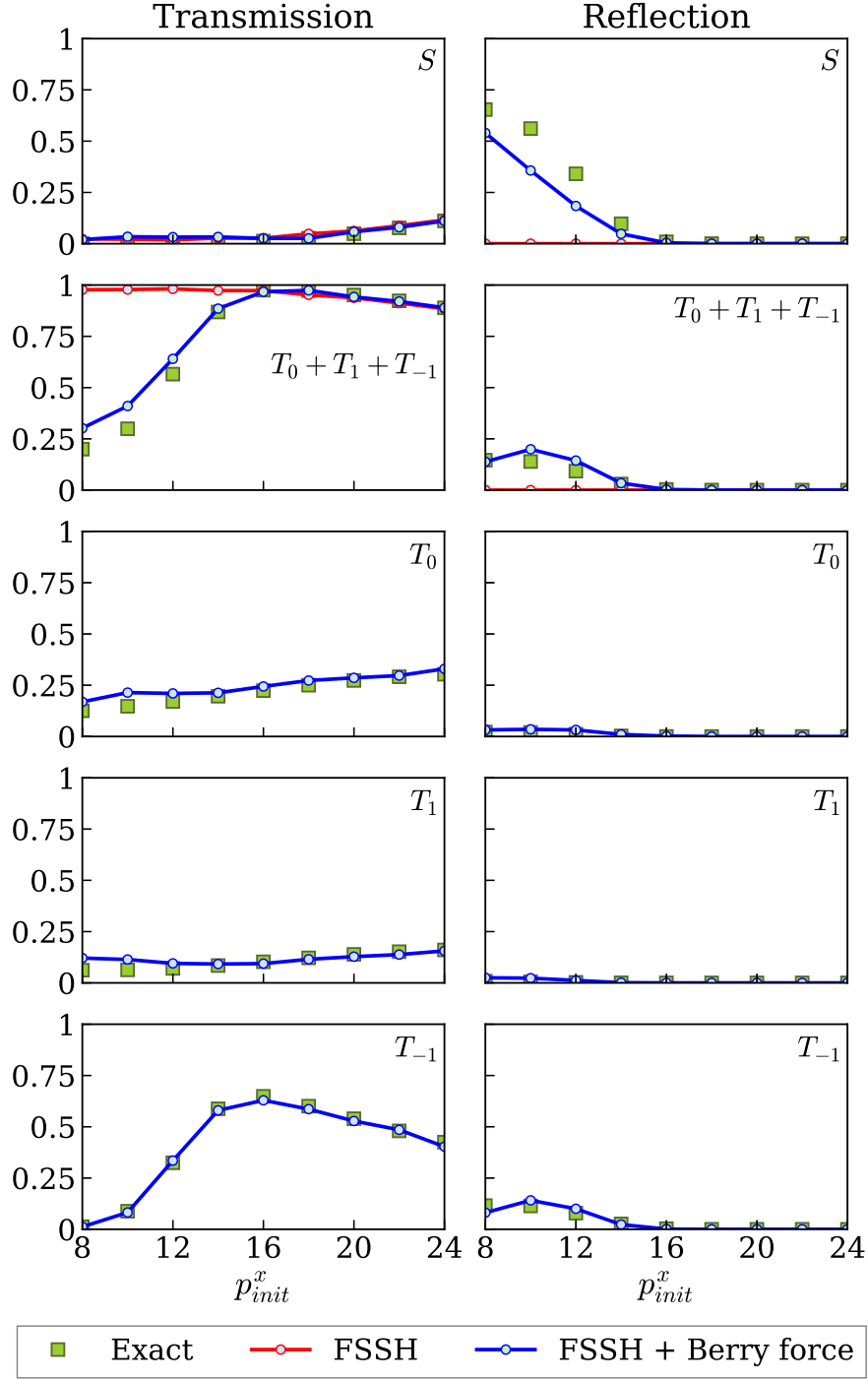}
    \caption{\label{fig:A010Sall} Probabilities for transmission and reflection on the upper surfaces, lower surfaces and each triplet spin-diabat as function of initial momentum $p^x_{init}$. The system is in the adiabatic limit, with $A  = 0.10$, $B = 3.0$ and $W = 5.0$; dynamics are initialized with $p^x_{init} = p^y_{init}$ on the upper singlet $\ket{S}$. The exact quantum dynamics data is compared with standard FSSH and our proposed FSSH with Berry force.  The upper transmission and lower reflection probabilities are calculated by summing over the three triplet diabatic states. Note that for each triplet diabat state, there is no data for standard FSSH because standard FSSH running on adiabats does not define an active diabat.  When system is initialized on the upper singlet state, standard FSSH predicts 100\% transmission but our approach can capture the reflection resulting by Berry force qualitatively.}
\end{figure}

We begin with the data where the wavepacket starts on the upper singlet $\ket{S}$ with $p^x_{init} = p^y_{init}$. The parameters for the model Hamiltonian are $A = 0.10$, $B = 3.0$ and $W = 5.0$.  Note that,   given this value of $A$, the system is largely in the adiabatic regime.
The transmission and reflection probabilities are presented in Fig. \ref{fig:A010Sall}.

According to Fig.\ref{fig:geometry}, even though the upper adiabatic surface is completely flat, the exact quantum results show that a huge amount of population is reflected when the momentum is low. The standard FSSH algorithm (without Berry force) fails and predicts zero reflection. 
By contrast, our new approach  can capture this Berry force induced reflection qualitatively (and nearly quantitatively). Not only do we recover strong estimates for reflection vs transmission, we also predict the population on each of the three degenerate triplet states fairly accurately.
 
In the supporting information, we provide data for a few different cases. In particular, we study:
\begin{enumerate}
    \item The case $A = 0.02$ -- which is the  nonadiabatic limit.
    \item The case that the system initialized on a lower singlet state (for both $A = 0.10$ and $A = 0.02$).
\end{enumerate} (We hold all other parameters are the same.)
For these problems, standard FSSH and Berry-force corrected FSSH yield fairly similar numerical results.

\subsection{Initialize on a triplet state}
\begin{figure} [H]
    \centering
    \includegraphics[width=0.75\columnwidth]{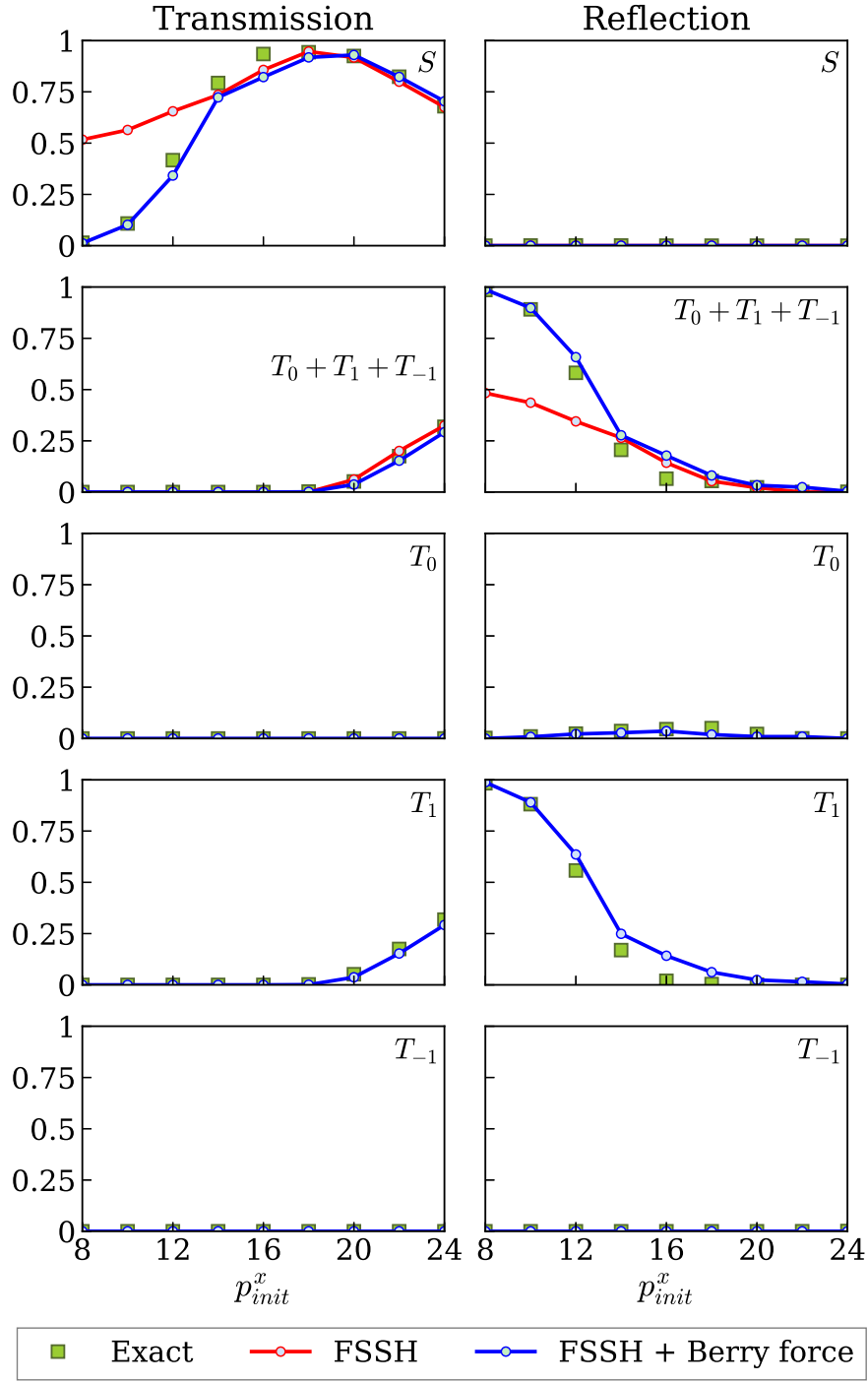}
    \caption{\label{fig:A0.10T1all} Probabilities for  transmission and reflection on the upper surfaces, lower surfaces and each triplet spin-diabat as a function of the initial momentum $p^x_{init}$. The system is in the adiabatic limit with $A  = 0.10$, $B = 3.0$ and $W = 5.0$ and initialized with $p^x_{init} = p^y_{init}$ on the lower triplet $\ket{T_1}$. 
    FSSH fails miserably at low momentum, whereas the new approach does well.}
\end{figure}

\begin{figure} [H]
    \centering
    \includegraphics[width=0.75\columnwidth]{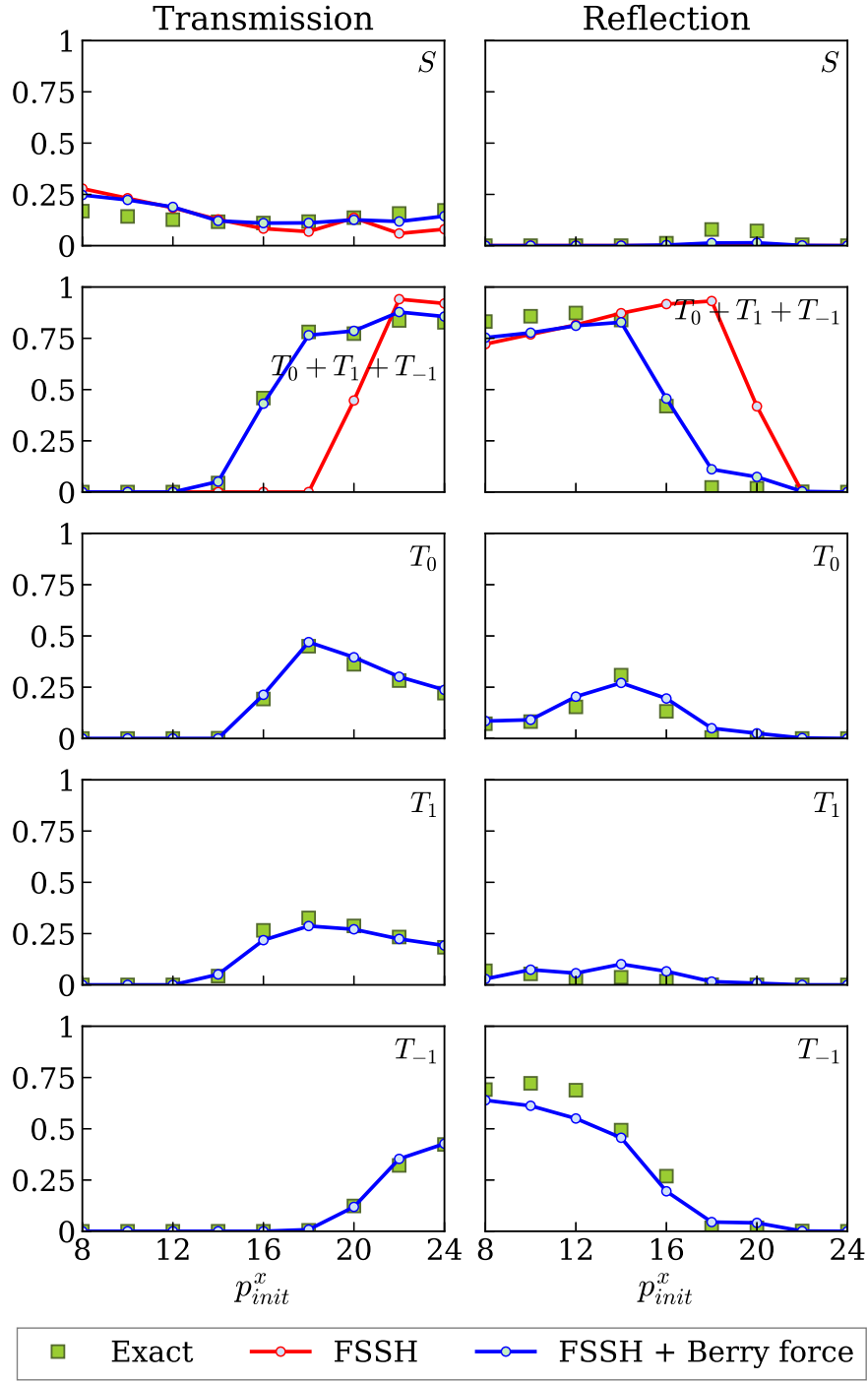}
    \caption{\label{fig:A0.10T-1all} Same as Fig. \ref{fig:A0.10T1all},  but now the system is initialized on  $\ket{T_{-1}}$. Again, the new method is more accurate compared to standard FSSH.}
\end{figure}

While the data above suggests that standard FSSH is sometimes good enough for dynamics initialized on a singlet, the case of initialization on a lower triplet is far more difficult, where there is competition between the
adiabatic force and the Berry force. To see this competition, consider three cases (with $W > 0$ and $A=0.10$):

\begin{enumerate} 
\item A wavepacket is initialized on $\ket{T_1}$.  Here, the
Berry force will create an energy barrier for transmission to both the upper and lower surfaces. 
As a result, perhaps not surprisingly, standard FSSH overestimates the transmission on the lower surface in Fig. \ref{fig:A0.10T1all}. 

\item A wavepacket is initialized on $\ket{T_{-1}}$.  Here, the Berry force will always decrease the energy barrier to transmit. Thus, it is not surprising that, in Fig. \ref{fig:A0.10T-1all}, standard FSSH underestimates the transmission on the upper surfaces.    

\item A wavepacket is initialized on $\ket{T_0}$.  Here, the Berry force can both increase and decrease the transmission barrier
depending on the electronic character of the different diabats. 
\end{enumerate}

In all three cases, our modified FSSH algorithm shows reasonable improvements in the population distribution  compared to standard FSSH.

\begin{figure} [H]
    \centering
    \includegraphics[width=0.75\columnwidth]{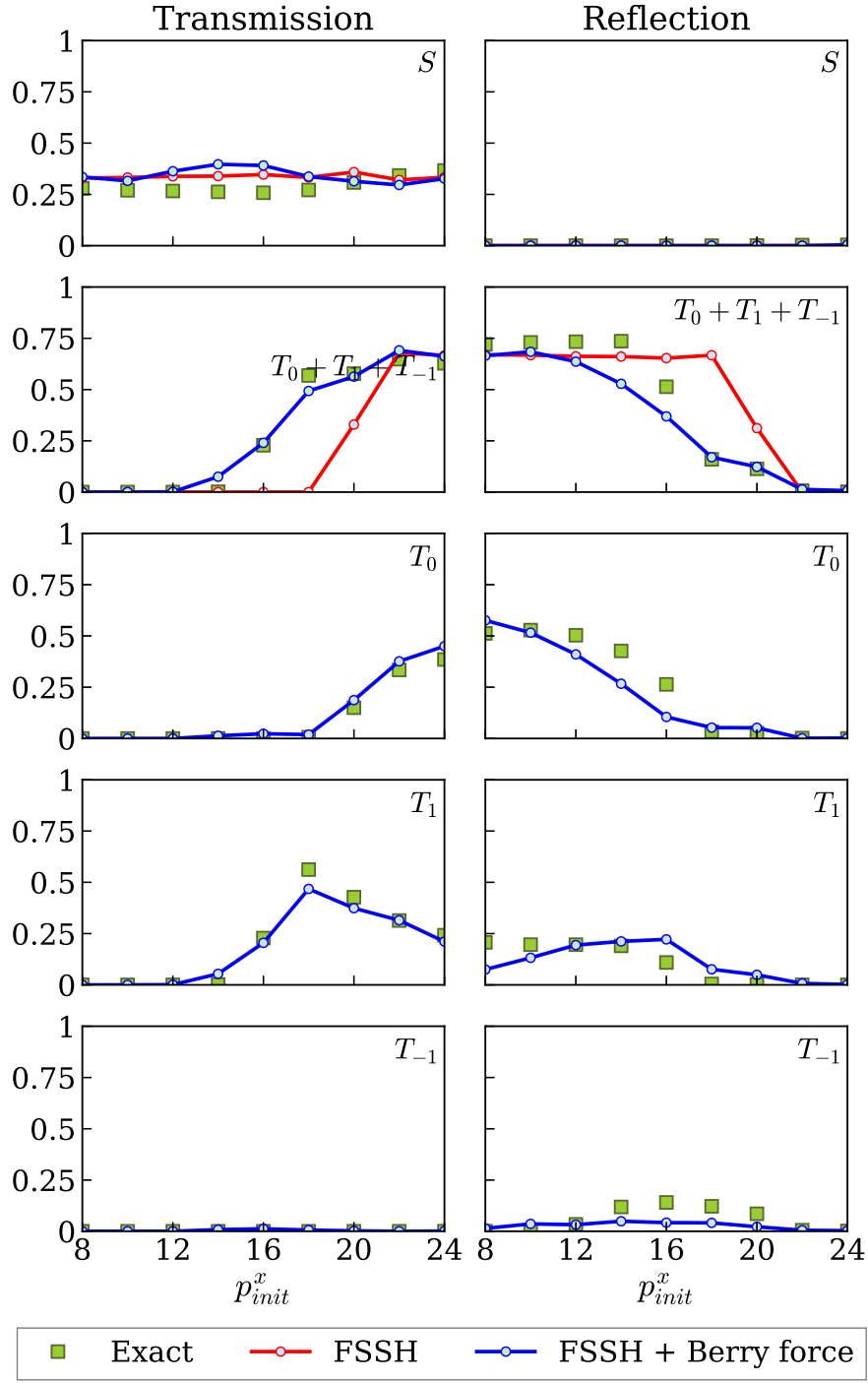}
    \caption{\label{fig:A0.10T0all} Same as Fig. \ref{fig:A0.10T1all},  but now the system is initialized on  $\ket{T_0}$. Again, our new method is better than standard FSSH.}
\end{figure}

\begin{figure} [H]
    \centering
    \includegraphics[width=0.75\columnwidth]{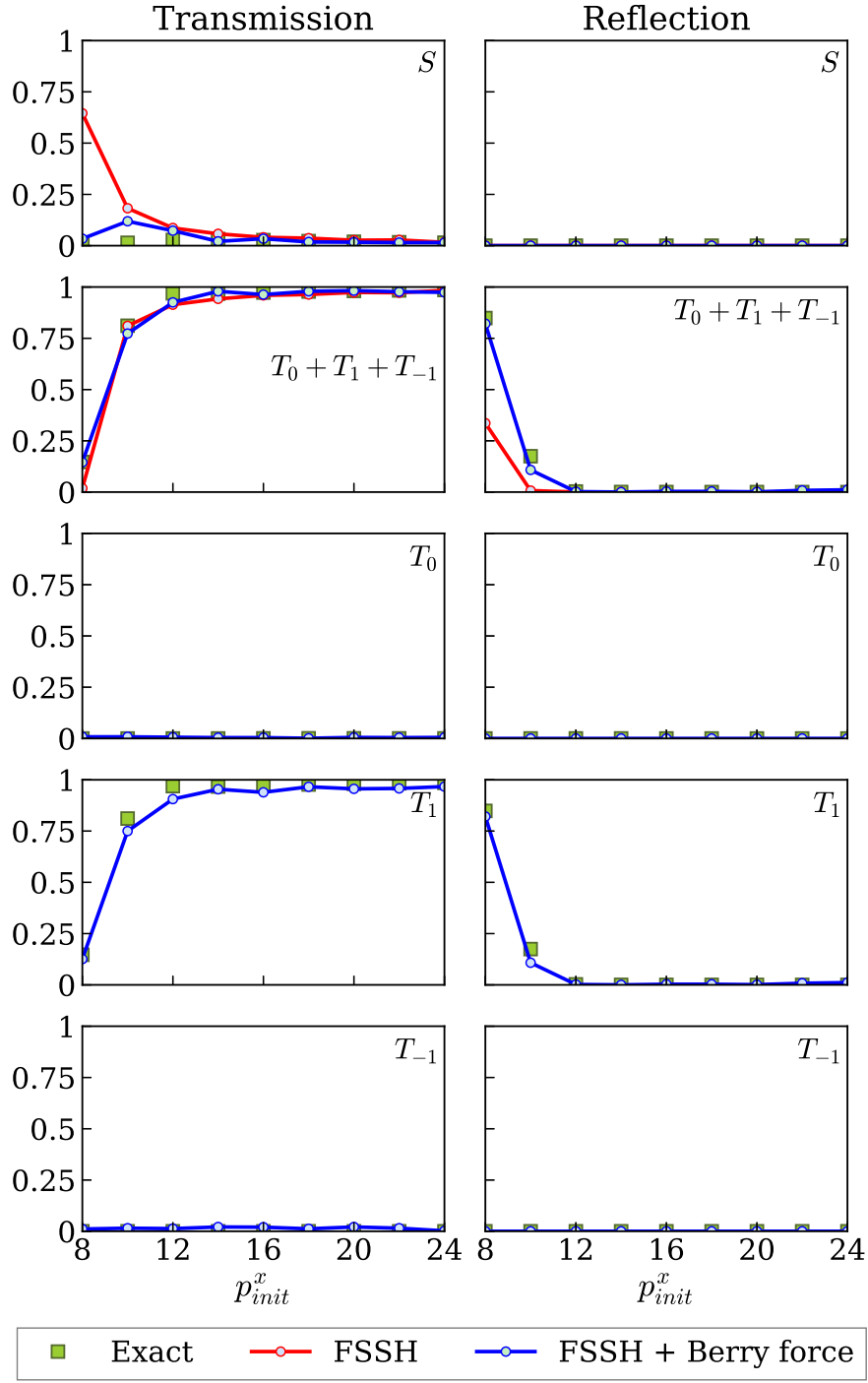}
    \caption{\label{fig:A0.02T1all} Data in the nonadiabatic limit. Here, we set all of the system parameters to be the same as in Fig. \ref{fig:A0.10T1all}, but we set $A = 0.02$ so as to operate in the nonadiabatic limit.}
\end{figure}

Lastly, let us consider the results above in the diabatic limit where $A = 0.02$ (and the system remains initialized on the lower triplet state).  Under such conditions, as shown in Fig. \ref{fig:A0.02T1all}, surface hopping with Berry force continues to perform better than normal FSSH.  For example, when a trajectory is initialized on the $\ket{T_{1}}$ state, the trajectory should reflect (because of the Berry force) if it continues towards the lowest adiabat (singlet). However, because standard FSSH lacks the Berry force, the algorithm overestimates the transmission on the singlet at low momentum. 
 
Note that, in Fig. \ref{fig:A0.02T1all}, we have used the criterion in Eq. \ref{eq:extreme} for establishing when to apply the berry force.  Without such a criterion for turning off the Berry force, one would find erroneous results for scattering with a large incoming momentum ($p_x = p_y > 10)$). Here the system is in the extreme diabatic limit (almost all trajectories will transit on the upper surface), and applying a local Berry force would lead to spurious reflection. 

In the end, assuming one is careful to always include such a cutoff for a Berry force, empirically we find that our results are always better (often much much better) than standard surface hopping. 

\mycomment{More numerical results, including data for the $A=0.02$ case with initialization on diabats $\ket{T_0}$ and $\ket{T_{-1}}$, can be found in the supporting information.}

In the end, the data above suggest that one can indeed use a modified FSSH code to treat degenerate nonadiabatic dynamics problems with complex-valued Hamiltonians and/or Berry forces.

\section{\label{sec:Discussion}Discussion}

\subsection{Momentum Rescaling and the Compatibility of the proposed quasi-diabatic Berry force with standard (two-state) non-degenerate Hamiltonian approaches}
 
For an electronic system with two states, the presence of a Berry force can explain most observations in the observed scattering simulations above. On the one hand, for trajectories that stay on the same adiabat,  the Berry curvatures of different 
adiabatic surfaces act as different magnetic fields, which eventually leads to a state specific momentum shift. 
On the other hand, for a trajectory that hops between adiabats and stays on the same diabat,  there is effectively zero contribution from a Berry force;
the entire concept can be safely ignored. 
This latter situation can be rationalized if we approximate that the transition between adiabats occurs exactly at the diabatic crossing point (say at position $x^{\ddagger}$).  In such a case, the 
trajectory will experience the Berry force on its initial adiabat from $x = -{\infty}$ until position  $x^{\ddagger}$ and then the equal and opposite Berry 
force on the other adiabat from $x^{\ddagger}$ to $x=\infty$. These two contributions cancel completely and such a zig-zag can actually be seen computationally (see Fig. 9 of Ref. \citenum{Miao2019}). 

Now, the situation above might seem to conform easily to (classical) FSSH dynamics. In practice, however, the FSSH algorithm allows independent trajectories to hop stochastically back and 
forth between adiabatic states an arbitrary number of times, and any given trajectory will experience an inconsistent Berry force (and therefore end up with wrong asymptotic momentum) if the trajectory does not necessarily hop at the crossing point or if it 
hops more than one time (again, not at the crossing point). As a result of the wrong momentum, if one applies FSSH normally, one will find incorrect nonadiabatic 
transmission and reflection probabilities. To solve this problem, in Ref. \citenum{Wu2021}, we 
suggested a dynamical rescaling momentum direction (i.e. a direction that depends on the nuclear momentum) which ensures there is a consistent momentum for each  adiabatic surface asymptotically. And overall, as shown in Ref. \citenum{Wu2021}, FSSH with an adiabatic Berry force and a dynamic rescaling direction can accurately simulate the complex two-state crossing problem. 

With this background in mind, one must wonder: is the four-state ISC model presented in Sec. \ref{sec:Methods} above consistent with the  two-state case from Ref. \citenum{Wu2021}?  For our ISC system with electronic degeneracy, the adiabats are not well-defined and
we have argued that a proper diabatic basis is needed to explain all the phenomena observed from 
wavepacket dynamics calculation; therefore, we have proposed here to use a quasi-diabatic Berry force, such that each trajectory is assigned a  nonzero Berry force. What is the relationship between 
%our different dynamical algorithm 
the present algorithm and the algorithm in Ref. \citenum{Wu2021}? 
%In general, we find that such a quasi-diabatic Berry force can be crucial for capturing accurate singlet-triplet crossing dynamics (at least for the Hamiltonian in Eq.\ref{eq:4stateHamiltonian}).
To that end, note that if, for a two-state model, we presume that quasi-diabats are unpaired (and come in a one-dimensional electronic subspace), then a quasi-diabat is clearly just an adiabat. In other words,  the quasi-diabatic Berry force in the present manuscript does reduce to the adiabatic Berry force in a two-state model.  The real difference between the two algorithms, however, is that in the former case, one must redefine quasi-diabatic energies on different sides of a crossing for stable FSSH dynamics (whereas in the latter, adiabatic energies can be defined globally).  In the end, we do believe that one can develop an FSSH model that can simulate both doublet-doublet and singlet-triplet crossings consistently.  
 
\subsection{The $\eta$ factor in the quasi-diabatic Berry force}
We next discuss the factor $\eta$ in our quasi-diabatic Berry force in Eq. \ref{eq:degeneratebf}. 
So far, we have argued for including such a  factor on purely phenomenological grounds--i.e. so that we can 
calculate a Berry force that leads to the correct momentum correction.  Practically speaking, we have shown that the algorithm  proposed above seems to apply to any singlet-triplet 
crossing which has the form of Eq. \ref{eq:4stateHamiltonian}. 

That being said, we will now show that the correct factor must depend on dimensionality.  To show this, consider the fictitious ``singlet-doublet'' 
crossing in Eq. \ref{eq:3stateHamiltonian} as an example
\begin{equation} \label{eq:3stateHamiltonian} H = A(x,y)\begin{pmatrix} \cos\theta    &  
    \frac {\sin\theta  e^{i\phi}} {\sqrt{2}}    &\frac { \sin\theta e^{-i\phi}} {\sqrt{2}}   \\
    \frac {\sin\theta e^{-i\phi}} {\sqrt{2}} &  -\cos\theta & 0 \\
    \frac {\sin\theta  e^{i\phi}}  {\sqrt{2}}  & 0 &  -\cos\theta 
\end{pmatrix} \end{equation}
Empirically, if one calculates derivative couplings, one will find that one must include a factor of $\eta = 1$ in order to achiever the right momentum correction. More generally, 
for a singlet crossing with $n$-fold degenerate states
\begin{equation} \label{eq:nstateHamiltonian} H = A(x,y)\begin{pmatrix} \cos\theta    &  \cdots & 
    \frac {\sin\theta  e^{i\phi}} {\sqrt{n}}    &\frac { \sin\theta e^{-i\phi}} {\sqrt{n}}   \\
    \vdots &  \ddots  & \vdots & \vdots &\\ 
    \frac {\sin\theta e^{-i\phi}} {\sqrt{n}} &\cdots  & -\cos\theta & 0 \\
    \frac {\sin\theta  e^{i\phi}}  {\sqrt{n}}  &  \cdots & 0 &  -\cos\theta 
\end{pmatrix} \end{equation}
one requires a factor $\eta = n/2$ (where $n$ is the dimensionality of degeneracy)
in order to achieve the correct asymptotic momenta.  For a general derivation of 
the $\eta$ factor as a function of the degeneracy dimensionality $n$, 
see the Appendix \ref{sec:Appendix}. 

\subsection{The limitations of our algorithm: Berry-Force Induced Tunneling}
\begin{figure} [H]
    \centering
    \includegraphics[width=0.75\columnwidth]{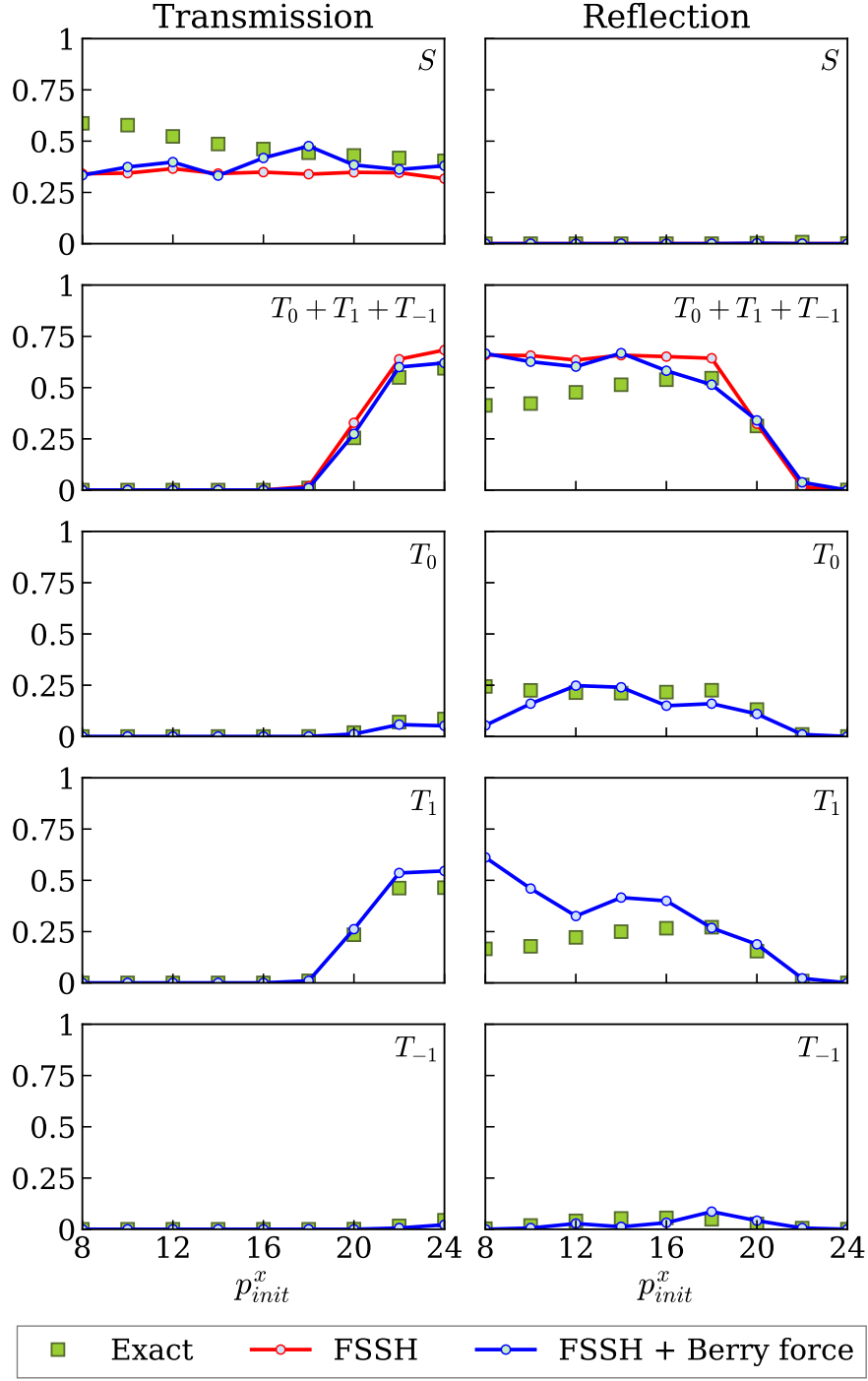}
    \caption{\label{fig:A010T1allang0} Tunneling failure of FSSH. We choose all system parameters to be the same as in Fig. \ref{fig:A0.10T1all} and the wavepacket is initialized on the lower triplet $\ket{T_1}$ coming in from $x=-\infty$. We set ${\vec p}_{init} = (p^x_{init} , 0)$.  We find FSSH cannot deliver quantitative accuracy and the trajectories cannot tunnel from triplet to  singlet.}

\end{figure}

Although we have so far presented encouraging data, we must now address one numerical example for which our modified FSSH algorithm does not deliver strong results.
This example is
the case of perpendicular incoming trajectories (${\vec p}_{init} = (p^x_{init} , 0)$) initialized on the lower triplet in the adiabatic limit ($A = 0.10$). For such a case,  the population distribution is shown in Fig. \ref{fig:A010T1allang0}. 
At low momentum, one can see that the surface hopping underestimates the transmission probability with or without a Berry force correction. In practice, including Berry force does not improve FSSH results. 

Why does this error emerge? Our preliminary answer is that this failure emerges
because of FSSH's inability to achieve tunneling: a trajectory entering from $x = -\infty$ along a triplet is repelled by the adiabatic force and simply cannot go far enough to positive $x$ in order to reach the crossing the region (and hop down to the singlet set). Thus, the failure of FSSH to include tunneling dooms the algorithm. Note that, interestingly, this tunneling feature can be accentuated when simulating complex-valued Hamiltonians with Berry forces. For instance, in the context of Fig. \ref{fig:A010T1allang0}, note that, if we set $W =0$, FSSH performs quite well.

\subsection{Open questions}

At this juncture, before concluding, we would like to highlight several open questions regarding the future of the present FSSH approach for simulating singlet-triplet crossing.
First, numerical results for the two-state model clearly suggest that one needs to incorporate decoherence within FSSH
to correctly treat Berry force induced wavepacket separation. \cite{Miao2019}
However, for the present model, it is not clear how important decoherence is;
nor is it clear how to include decoherence in our degenerate model.
\mycomment{For problems without degeneracy (e.g. real-valued two state problems), it is now widely established\cite{Bittner1995, Jasper2005, Subotnik2011, Kapral2016} that the correct decoherence rate must be proportional to the Born-Oppenheimer force difference between adiabats, $\Delta \vec F = \vec F_1 - \vec F_2$. But is this still true if pseudo-magnetic fields are present? Should the pseudomagnetic fields also be included in the direction $\Delta \vec F$  above? The answer to date is unknown. Moreover, 
as far as a decoherence procedure, the usual approach is to} collapse a trajectory's electronic amplitude to 
one adiabatic surface so to avoid incorrect hops in the future. Is this still valid for our quasi-diabatic scheme? In general, how should the quasi-diabat be taken into account when determining the final state after a decoherence event for  multidimensional  system with Berry forces?  These questions must be addressed in the future. So far, no decoherence has been included in Sec. \ref{sec:Results}. 

Second, again on the practical side, we found above that reversing a trajectory's momenta after a frustrated quasi-diabatic hop can sometimes help (but other can times can hurt) as far recovering the correct transmission reflection ratio and momentum. In order to account for this observation, for the nearly flat model Hamiltonian in Eq. \ref{eq:4stateHamiltonian}, we have proposed that one should implement momentum reversal rule only for the cases that satisfy Eq. \ref{eq:reverse} -- and that has been very helpful so far. More generally, however, more research into momentum reversal and the optimal protocol will be essential in the future, where we will necessarily need to investigate model Hamiltonians that are not flat and models with more than four electronic states. 

Third, for our FSSH purposes in this paper, we have required the existence of a ``proper'' diabatic basis.  In the context of  Eq. \ref{eq:4stateHamiltonian}, 
such a proper diabatic must  satisfy two requirements.  (a) These diabats must reduce to adiabatic states outside of any coupling region. (b) Even more importantly, for the triplet diabats, if one investigates how their coupling to the singlet changes with position, one requires that the couplings should be as localized as possible in momentum space.  For instance, for the crossings chosen in this paper, the couplings from $\ket{T_1}$ or $\ket{T_{-1}}$ to the singlet vary as $e^{\pm i\phi}$ -- a choice which leads to very different Berry forces on the different quasi-diabats. More generally, however, one must wonder: are these requirements general? Are they applicable and practical for more complicated Hamiltonians?  Is it practical to find such diabats in an {\em ab initio} simulation? Is it reasonable to assume that, for a realistic Hamiltonian, one can find diabats for which the direction of the gradient of the diabatic energy ($x$) and the gradient of the phase of the diabatic coupling ($y$) do not change appreciably over the course of a crossing? The present algorithm is not yet general and will need to be further developed in the future. \mycomment{Obviously, for an {\em ab initio} simulation, the goal will be to propagate dynamics with as little information as possible about the potential energy surfaces (and certainly including only local information).}

Fourth and finally, in this paper, we have not addressed (at all) the possibility of a true, externally generated magnetic field. In such a case, no triplet spin-adiabats will 
be degenerate and each spin will be partially aligned with the magnetic field. Now, in the limit of a very very large magnetic field, presumably, one splits all electronic states and the problem reduces to a standard (non-degenerate) nonadiabatic problem. Thus, one must wonder: if the magnetic field is large enough,   do we still need Berry forces and a quasi-diabatic basis? Or can we simply use normal FSSH with an adiabatic basis? 
More generally: can we use FSSH to learn how the 
direction and strength of an external magnetic field will influence the spin-dependent dynamics of ISC processes?   Might such effects be relevant for modeling the molecular dynamics underlying bird navigation with magnetoreception \cite{ Mouritsen2018, Kerpal2019} and/or magnetic field effects in organic photochemistry \cite{Steiner1989, Hore2020, Higgins2020}?

\section{\label{sec:Summary}Summary}
In conclusion, we have proposed an extension of FSSH to include Berry forces within a degenerate singlet-triplet crossing 
model. With our modified algorithm, FSSH can capture most of the important features of a scattering process -- the probabilities of transmission and reflection
and the scattering momentum-shift.  The present algorithm is not yet general to arbitrary Hamiltonians, but the present manuscript highlights a new perspective on how to semiclassically treat the case of nonadiabatic dynamics with strong electronic degeneracy.  In the future, merging the domains of nonadiabatic dynamics and spintronics offers many new and exciting areas for research, and we hope that the present semiclassical perspective will help us explore this vast new terrain.

\appendix
 
\section{\label{sec:Appendix} Proof of the factor in quasi-diabatic Berry force expression} 
Below, we will show how to construct the factor $\eta$  that is required for the quasi-diabatic Berry force expression in Eq.  
\ref{eq:degeneratebf}. We will show that such a factor depends critically on the dimensionality of the degenerate space, and for the singlet-triplet problem, the factor should be $\eta = 3/2$.

Let us consider a general model of $n$-fold degenerate multiplets crossing with one singlet state. (Above, we have considered $n=3$, the case of a triplet.) For the sake of simplicity, we 
assume that the coupling between the singlet and each multiplet has the same magnitude, even though each coupling can carry a different complex phase. 

\begin{equation}
    \label{eq:NstateHamiltonian}
    H= \begin{pmatrix}
     \epsilon & V e^{i\phi_1} & \cdots & Ve^{i\phi_n} \\
     V e^{-i\phi_1} & -\epsilon & 0 & 0 \\
     \vdots  & 0 & \ddots & 0 \\ 
     V e^{-i\phi_n} & 0 & 0 & -\epsilon \\
    \end{pmatrix}
\end{equation}

For simplicity of notation, the adiabatic basis $\{\psi\}$ is indexed by $\{ijk\}$ and the original diabatic basis in Eq. \ref{eq:NstateHamiltonian} $\{\phi\}$ is indexed by $\{abc\}$. Obviously the quasi-diabatic basis $\{\tilde \phi \}$ can also be indexed by $\{abc\}$.

By analogy with problems in electrodynamics, it is easy to show that 
there will be two distinct adiabats at low and high energy;  we shall call these states the upper and lower 
polaritons (even though there is no light involved). There remain $n-1$ degenerate adiabats with intermediate energy.
Again, by analogy to electrodynamics, we will rotate the $n$ degenerate diabats into a new basis,  which is composed of one ``bright'' state ($\ket{B}$, that is  coupled to the singlet) and 
$n-1$ ``dark'' states $(\left\{ \ket{D_{\alpha}} \right\}$, which do not couple directly to the singlet [but which do coupled to the bright state]).

For our model Hamiltonian in Eq. \ref{eq:NstateHamiltonian},
both the two polaritonic states $\ket{\psi_{UP}}$,$\ket{\psi_{LP}}$ and the bright state $\ket{B}$ can be written down analytically
\begin{equation} \ket{\psi_{UP}} = \frac 1 {\sqrt{2(\epsilon^2 + nV^2 + \epsilon\sqrt{\epsilon^2 + nV^2})}} \left(\epsilon + \sqrt{\epsilon + nV^2}, Ve^{-i\phi_1},\cdots, Ve^{-i\phi_n}  \right)^T \end{equation}
\begin{equation} \label{eq:LP}\ket{\psi_{LP}} = \frac 1 {\sqrt{2(\epsilon^2 + nV^2 - \epsilon\sqrt{\epsilon^2 + nV^2})}} \left(\epsilon - \sqrt{\epsilon + nV^2}, Ve^{-i\phi_1},\cdots, Ve^{-i\phi_n}  \right)^T \end{equation}
\begin{equation} \label{eq:B}
    \ket{B} = \frac 1 {\sqrt{n}} (0, e^{-i\phi_1}, \cdots, e^{-i\phi_n})^T
\end{equation}

Then, we isolate the $n$ adiabats closest to the original $n$-fold degenerate multiplets and rotate the adiabats (with rotation matrix $\bm R$) to align them with the diabats; the result is a set of degenerate quasi-diabats $\{\tilde \phi\}$ defined by $ \ket{\tilde \phi_a} = \sum_{i} \ket{\psi_i} R_{ia}$.
The optimized rotation matrix satisfies the Kabsch algorithm and can be solved by $\bm R = \bm O^\dagger (\bm O \bm O^\dagger)^{-\frac 1 2} $ where $O_{ai} =\bra{\phi_a} \ket{\psi_i}$.  
The quasi-diabats can therefore be expressed as:
\begin{equation}\begin{aligned} \label{eq:etaa}
    \ket{\tilde \phi_a} &= \sum_{i} \ket{\psi_i} R_{ia}  = \sum_{i,a'} \ket{\psi_i} \bra{\psi_i} \ket{\phi_{a'}} \bra{\phi_{a'}} \bigg(\hat P_M \sum_j \ket{\psi_j}\bra{\psi_j}\hat P_M\bigg)^{-\frac 1 2} \ket{\phi}\\
\end{aligned}\end{equation}
where $\hat P_M = \sum_a \ket{\phi_a}\bra{\phi_a} = \ket{B}\bra{B} +  \sum_{\alpha} \ket{D_{\alpha}}\bra{D_{\alpha}}$ is the projector onto the multiplet subspace. Without loss of generality, consider the case where we rotate
the upper $n$  adiabats to generate quasi-diabats (so that we do not include  $\ket{\psi_{LP}}$ in our projection $\sum_j \ket{\psi_j}\bra{\psi_j}$ in Eq. \ref{eq:etaa}). In such a case, the projector to the upper $n$ adiabats can is simply:  
$\sum_j \ket{\psi_j}\bra{\psi_j} = \hat I - \ket{\psi_{LP}}\bra{\psi_{LP}}$. 

Finally, since the multiplet manifold couples to the singlet only through the bright state, the projection of the lower polariton to the multiplet subspace has a contribution only from the bright state:
\begin{equation} \hat P_M \ket{\psi_{LP}} \bra{\psi_{LP}} \hat P_M = \abs{\bra{B}\ket{\psi_{LP}}}^2 \ket{B} \bra{B}  \end{equation}
Therefore, we can rewrite Eq. \ref{eq:etaa} as 
\begin{equation}\begin{aligned}  
    \ket{\tilde \phi_a}  &= \sum_{a'} \bigg( \hat I - \ket{\psi_{LP}}\bra{\psi_{LP}} \bigg)\ket{\phi_{a'}} \bra{\phi_{a'}} \bigg( \hat P_M  \bigg( \hat I - \ket{\psi_{LP}}\bra{\psi_{LP}} \bigg) \hat P_M\bigg)^{-\frac 1 2} \ket{\phi_a}\\
    &=   \sum_{a'} \bigg( \hat I - \ket{\psi_{LP}}\bra{\psi_{LP}} \bigg)\ket{\phi_{a'}} \bra{\phi_{a'}} \bigg(  (1 - \abs{\bra{B}\ket{\psi_{LP}}}^2)^{-\frac 1 2} \ket{B} \bra{B}  + \sum_\alpha \ket{D_{\alpha}}\bra{D_{\alpha}} \bigg)  \ket{\phi_a} \\
    &=   \sum_{a'} \bigg( \hat I - \ket{\psi_{LP}}\bra{\psi_{LP}} \bigg)\ket{\phi_{a'}} \bra{\phi_{a'}} \bigg(  (1 - \abs{\bra{B}\ket{\psi_{LP}}}^2)^{-\frac 1 2} \ket{B} \bra{B}  + 
    \left(\hat{I} - \ket{B}\bra{B} \right) \bigg)  \ket{\phi_a} \\
    &= \sum_{a'}  (\ket{\phi_{a'}} - \bra{\psi_{LP}}\ket{\phi_{a'}}\ket{\psi_{LP}} ) \left(  \delta_{aa'} + ((1- \abs{\bra{B} \ket{\psi_{LP}}}^2)^{-\frac 1 2} - 1) \bra{\phi_{a'}} \ket{B}\bra{B}\ket{\phi_{a}} \right)  \\
\end{aligned}\end{equation}

The quasi-diabat expression above can be evaluated analytically by inserting Eq. \ref{eq:LP} and Eq. \ref{eq:B}
\begin{equation}\begin{aligned} \label{eq:etaai}
    \ket{\tilde \phi_a}_i  = \delta_{ai} + \left(\left( \sqrt{ \frac{\Delta^2} {\Delta^2 - nV^2}  }- 1\right) \left(1 - \frac {nV^2} {\Delta^2}\right) - \frac {nV^2} {\Delta^2} \right) \frac {e^{-i(\phi_{i}-\phi_a)}} {n}
\end{aligned}\end{equation}  
where $\Delta = {\sqrt{2(\epsilon^2 + nV^2 - \epsilon\sqrt{\epsilon^2 + nV^2})}}$.

At this point, let us suppose that the Hamiltonian has only two nuclear degrees of freedom, such that $\epsilon$ and $V$ depend on $\hat x$ and $\phi$ depends on $\hat y$.     
We can heuristically write Eq. \ref{eq:etaai} as $ \ket{\tilde \phi_a}_i  = \delta_{ai} + f(x) \frac {e^{-i(\phi_{i}-\phi_a)}} {n} $. Then, the derivative coupling and the 
quasi-diabatic Berry curvature are  
\begin{equation} {d}_{aa}^x = \bra{\tilde \phi_a} \nabla_x \ket{\tilde \phi_a} =  \frac 1 {n^2} \frac {\partial f(x)} {\partial x} f(x) \end{equation}
\begin{equation} \label{eq:dy} { d}_{aa}^y = \bra{\tilde \phi_a} \nabla_y \ket{\tilde \phi_a} = \sum_{i} -\frac i {n^2} \frac {\partial (\phi_i - \phi_a)} {\partial y} f^2(x) =  -\frac i {n} \frac {\partial \phi_a}  {\partial y} f^2(x) \end{equation}
\begin{equation}  \label{eq:omegaxy} { \Omega}_{a}^{xy} = i (\nabla_x \bra{\tilde \phi_a} \nabla_y \ket{\tilde \phi_a} - \nabla_y \bra{\tilde \phi_a} \nabla_x \ket{\tilde \phi_a}) =   \frac 2 {n} \frac {\partial \phi_a}  {\partial y} \frac {\partial f(x)} {\partial x} f(x)  \end{equation}
Here in Eq. \ref{eq:dy}, we have assumed that the sum of the derivatives all phase factors is zero, i.e. $\sum_i \frac {\partial \phi_i} {\partial y} = 0$; such an assumption is clearly valid for the singlet-triplet case without a magnetic field. 

In the end,  note that the quasi-diabatic Berry curvature in Eq. \ref{eq:omegaxy} is inversely proportional to the dimensionality of the degenerate multiplet manifold.  Thus, in general, in order to reach the correct asymptotic momentum, we must multiply each quasi-diabatic Berry force by a factor of $n/2$. \cite{footnote3} 

\begin{suppinfo}

Additional numerical results can be found in the supporting information, including the cases: (1) $A = 0.02$ and the system is initialized on the upper singlet state; (2)  the system is initialized on the lower singlet state (for both $A = 0.10$ and $A = 0.02$)  (3)  $A=0.02$ and the system is initialized on diabat $\ket{T_0}$ or diabat $\ket{T_{-1}}$.
\end{suppinfo}
 
\bibliography{ref.bib}

\providecommand{\latin}[1]{#1}
\makeatletter
\providecommand{\doi}
  {\begingroup\let\do\@makeother\dospecials
  \catcode`\{=1 \catcode`\}=2 \doi@aux}
\providecommand{\doi@aux}[1]{\endgroup\texttt{#1}}
\makeatother
\providecommand*\mcitethebibliography{\thebibliography}
\csname @ifundefined\endcsname{endmcitethebibliography}
  {\let\endmcitethebibliography\endthebibliography}{}
\begin{mcitethebibliography}{55}
\providecommand*\natexlab[1]{#1}
\providecommand*\mciteSetBstSublistMode[1]{}
\providecommand*\mciteSetBstMaxWidthForm[2]{}
\providecommand*\mciteBstWouldAddEndPuncttrue
  {\def\EndOfBibitem{\unskip.}}
\providecommand*\mciteBstWouldAddEndPunctfalse
  {\let\EndOfBibitem\relax}
\providecommand*\mciteSetBstMidEndSepPunct[3]{}
\providecommand*\mciteSetBstSublistLabelBeginEnd[3]{}
\providecommand*\EndOfBibitem{}
\mciteSetBstSublistMode{f}
\mciteSetBstMaxWidthForm{subitem}{(\alph{mcitesubitemcount})}
\mciteSetBstSublistLabelBeginEnd
  {\mcitemaxwidthsubitemform\space}
  {\relax}
  {\relax}

\bibitem[Tully(1990)]{Tully1990}
Tully,~J.~C. Molecular dynamics with electronic transitions. \emph{The Journal
  of Chemical Physics} \textbf{1990}, \emph{93}, 1061\relax
\mciteBstWouldAddEndPuncttrue
\mciteSetBstMidEndSepPunct{\mcitedefaultmidpunct}
{\mcitedefaultendpunct}{\mcitedefaultseppunct}\relax
\EndOfBibitem
\bibitem[Tully(1998)]{Tully1998}
Tully,~J.~C. Mixed quantum–classical dynamics. \emph{Faraday Discussions}
  \textbf{1998}, \emph{110}, 407--419\relax
\mciteBstWouldAddEndPuncttrue
\mciteSetBstMidEndSepPunct{\mcitedefaultmidpunct}
{\mcitedefaultendpunct}{\mcitedefaultseppunct}\relax
\EndOfBibitem
\bibitem[Barbatti(2011)]{Barbatti2011}
Barbatti,~M. Nonadiabatic dynamics with trajectory surface hopping method.
  \emph{Wiley Interdisciplinary Reviews: Computational Molecular Science}
  \textbf{2011}, \emph{1}, 620--633\relax
\mciteBstWouldAddEndPuncttrue
\mciteSetBstMidEndSepPunct{\mcitedefaultmidpunct}
{\mcitedefaultendpunct}{\mcitedefaultseppunct}\relax
\EndOfBibitem
\bibitem[Mai \latin{et~al.}(2018)Mai, Marquetand, and González]{Mai2018}
Mai,~S.; Marquetand,~P.; González,~L. Nonadiabatic dynamics: The SHARC
  approach. \emph{Wiley Interdisciplinary Reviews: Computational Molecular
  Science} \textbf{2018}, \emph{8}, e1370\relax
\mciteBstWouldAddEndPuncttrue
\mciteSetBstMidEndSepPunct{\mcitedefaultmidpunct}
{\mcitedefaultendpunct}{\mcitedefaultseppunct}\relax
\EndOfBibitem
\bibitem[Nelson \latin{et~al.}(2020)Nelson, White, Bjorgaard, Sifain, Zhang,
  Nebgen, Fernandez-Alberti, Mozyrsky, Roitberg, and Tretiak]{Nelson2020}
Nelson,~T.~R.; White,~A.~J.; Bjorgaard,~J.~A.; Sifain,~A.~E.; Zhang,~Y.;
  Nebgen,~B.; Fernandez-Alberti,~S.; Mozyrsky,~D.; Roitberg,~A.~E.; Tretiak,~S.
  Non-adiabatic Excited-State Molecular Dynamics: Theory and Applications for
  Modeling Photophysics in Extended Molecular Materials. \emph{Chemical
  Reviews} \textbf{2020}, \emph{120}, 2215--2287\relax
\mciteBstWouldAddEndPuncttrue
\mciteSetBstMidEndSepPunct{\mcitedefaultmidpunct}
{\mcitedefaultendpunct}{\mcitedefaultseppunct}\relax
\EndOfBibitem
\bibitem[Kapral(2016)]{Kapral2016}
Kapral,~R. Surface hopping from the perspective of quantum--classical Liouville
  dynamics. \emph{Chemical Physics} \textbf{2016}, \emph{481}, 77--83\relax
\mciteBstWouldAddEndPuncttrue
\mciteSetBstMidEndSepPunct{\mcitedefaultmidpunct}
{\mcitedefaultendpunct}{\mcitedefaultseppunct}\relax
\EndOfBibitem
\bibitem[Bittner and Rossky(1995)Bittner, and Rossky]{Bittner1995}
Bittner,~E.~R.; Rossky,~P.~J. Quantum decoherence in mixed quantum‐classical
  systems: Nonadiabatic processes. \emph{The Journal of Chemical Physics}
  \textbf{1995}, \emph{103}, 8130\relax
\mciteBstWouldAddEndPuncttrue
\mciteSetBstMidEndSepPunct{\mcitedefaultmidpunct}
{\mcitedefaultendpunct}{\mcitedefaultseppunct}\relax
\EndOfBibitem
\bibitem[Jasper and Truhlar(2005)Jasper, and Truhlar]{Jasper2005}
Jasper,~A.~W.; Truhlar,~D.~G. Electronic decoherence time for
  non-Born-Oppenheimer trajectories. \emph{The Journal of Chemical Physics}
  \textbf{2005}, \emph{123}, 064103\relax
\mciteBstWouldAddEndPuncttrue
\mciteSetBstMidEndSepPunct{\mcitedefaultmidpunct}
{\mcitedefaultendpunct}{\mcitedefaultseppunct}\relax
\EndOfBibitem
\bibitem[Subotnik and Shenvi(2011)Subotnik, and Shenvi]{Subotnik2011}
Subotnik,~J.~E.; Shenvi,~N. A new approach to decoherence and momentum
  rescaling in the surface hopping algorithm. \emph{The Journal of Chemical
  Physics} \textbf{2011}, \emph{134}, 024105\relax
\mciteBstWouldAddEndPuncttrue
\mciteSetBstMidEndSepPunct{\mcitedefaultmidpunct}
{\mcitedefaultendpunct}{\mcitedefaultseppunct}\relax
\EndOfBibitem
\bibitem[Subotnik \latin{et~al.}(2016)Subotnik, Jain, Landry, Petit, Ouyang,
  and Bellonzi]{Subotnik2016}
Subotnik,~J.~E.; Jain,~A.; Landry,~B.; Petit,~A.; Ouyang,~W.; Bellonzi,~N.
  Understanding the Surface Hopping View of Electronic Transitions and
  Decoherence. \emph{Annual Review of Physical Chemistry} \textbf{2016},
  \emph{67}, 387--417\relax
\mciteBstWouldAddEndPuncttrue
\mciteSetBstMidEndSepPunct{\mcitedefaultmidpunct}
{\mcitedefaultendpunct}{\mcitedefaultseppunct}\relax
\EndOfBibitem
\bibitem[Jasper \latin{et~al.}(2001)Jasper, Hack, and Truhlar]{Jasper2001}
Jasper,~A.~W.; Hack,~M.~D.; Truhlar,~D.~G. The treatment of classically
  forbidden electronic transitions in semiclassical trajectory surface hopping
  calculations. \emph{The Journal of Chemical Physics} \textbf{2001},
  \emph{115}, 1804\relax
\mciteBstWouldAddEndPuncttrue
\mciteSetBstMidEndSepPunct{\mcitedefaultmidpunct}
{\mcitedefaultendpunct}{\mcitedefaultseppunct}\relax
\EndOfBibitem
\bibitem[Jasper and Truhlar(2003)Jasper, and Truhlar]{Jasper2003}
Jasper,~A.~W.; Truhlar,~D.~G. Improved treatment of momentum at classically
  forbidden electronic transitions in trajectory surface hopping calculations.
  \emph{Chemical Physics Letters} \textbf{2003}, \emph{369}, 60--67\relax
\mciteBstWouldAddEndPuncttrue
\mciteSetBstMidEndSepPunct{\mcitedefaultmidpunct}
{\mcitedefaultendpunct}{\mcitedefaultseppunct}\relax
\EndOfBibitem
\bibitem[Schwartz and Rossky(1994)Schwartz, and Rossky]{Schwartz1994}
Schwartz,~B.~J.; Rossky,~P.~J. Aqueous solvation dynamics with a quantum
  mechanical Solute: Computer simulation studies of the photoexcited hydrated
  electron. \emph{J. Chem. Phys} \textbf{1994}, \emph{101}, 6902\relax
\mciteBstWouldAddEndPuncttrue
\mciteSetBstMidEndSepPunct{\mcitedefaultmidpunct}
{\mcitedefaultendpunct}{\mcitedefaultseppunct}\relax
\EndOfBibitem
\bibitem[Nelson \latin{et~al.}(2011)Nelson, Fernandez-Alberti, Chernyak,
  Roitberg, and Tretiak]{Nelson2011}
Nelson,~T.; Fernandez-Alberti,~S.; Chernyak,~V.; Roitberg,~A.~E.; Tretiak,~S.
  Nonadiabatic Excited-State Molecular Dynamics Modeling of Photoinduced
  Dynamics in Conjugated Molecules. \emph{Journal of Physical Chemistry B}
  \textbf{2011}, \emph{115}, 5402--5414\relax
\mciteBstWouldAddEndPuncttrue
\mciteSetBstMidEndSepPunct{\mcitedefaultmidpunct}
{\mcitedefaultendpunct}{\mcitedefaultseppunct}\relax
\EndOfBibitem
\bibitem[Zaari and Varganov(2015)Zaari, and Varganov]{Zaari2015}
Zaari,~R.~R.; Varganov,~S.~A. Nonadiabatic transition state theory and
  trajectory surface hopping dynamics: intersystem crossing between 3B1 and 1A1
  states of SiH2. \emph{The Journal of Physical Chemistry A} \textbf{2015},
  \emph{119}, 1332--1338\relax
\mciteBstWouldAddEndPuncttrue
\mciteSetBstMidEndSepPunct{\mcitedefaultmidpunct}
{\mcitedefaultendpunct}{\mcitedefaultseppunct}\relax
\EndOfBibitem
\bibitem[Chakraborty \latin{et~al.}(2020)Chakraborty, Liu, Weinacht, and
  Matsika]{Chakraborty2020}
Chakraborty,~P.; Liu,~Y.; Weinacht,~T.; Matsika,~S. Excited state dynamics of
  cis, cis-1, 3-cyclooctadiene: Non-adiabatic trajectory surface hopping.
  \emph{The Journal of chemical physics} \textbf{2020}, \emph{152},
  174302\relax
\mciteBstWouldAddEndPuncttrue
\mciteSetBstMidEndSepPunct{\mcitedefaultmidpunct}
{\mcitedefaultendpunct}{\mcitedefaultseppunct}\relax
\EndOfBibitem
\bibitem[Landry and Subotnik(2014)Landry, and Subotnik]{Landry2014}
Landry,~B.~R.; Subotnik,~J.~E. Quantifying the Lifetime of Triplet Energy
  Transfer Processes in Organic Chromophores: A Case Study of
  4-(2-Naphthylmethyl)benzaldehyde. \emph{Journal of Chemical Theory and
  Computation} \textbf{2014}, \emph{10}, 4253--4263\relax
\mciteBstWouldAddEndPuncttrue
\mciteSetBstMidEndSepPunct{\mcitedefaultmidpunct}
{\mcitedefaultendpunct}{\mcitedefaultseppunct}\relax
\EndOfBibitem
\bibitem[Atkins and Gonzaíez(2017)Atkins, and Gonzaíez]{Atkins2017}
Atkins,~A.~J.; Gonzaíez,~L. Trajectory Surface-Hopping Dynamics Including
  Intersystem Crossing in [Ru(bpy) 3 ] 2+. \emph{J. Phys. Chem. Lett}
  \textbf{2017}, \emph{8}, 58\relax
\mciteBstWouldAddEndPuncttrue
\mciteSetBstMidEndSepPunct{\mcitedefaultmidpunct}
{\mcitedefaultendpunct}{\mcitedefaultseppunct}\relax
\EndOfBibitem
\bibitem[Shenvi \latin{et~al.}(2009)Shenvi, Roy, and Tully]{Shenvi2009}
Shenvi,~N.; Roy,~S.; Tully,~J.~C. Dynamical steering and electronic excitation
  in NO scattering from a gold surface. \emph{Science} \textbf{2009},
  \emph{326}, 829--832\relax
\mciteBstWouldAddEndPuncttrue
\mciteSetBstMidEndSepPunct{\mcitedefaultmidpunct}
{\mcitedefaultendpunct}{\mcitedefaultseppunct}\relax
\EndOfBibitem
\bibitem[Golibrzuch \latin{et~al.}(2014)Golibrzuch, Shirhatti, Rahinov,
  Kandratsenka, Auerbach, Wodtke, and Bartels]{Golibrzuch2014}
Golibrzuch,~K.; Shirhatti,~P.~R.; Rahinov,~I.; Kandratsenka,~A.;
  Auerbach,~D.~J.; Wodtke,~A.~M.; Bartels,~C. The importance of accurate
  adiabatic interaction potentials for the correct description of
  electronically nonadiabatic vibrational energy transfer: A combined
  experimental and theoretical study of NO(v = 3) collisions with a Au(111)
  surface. \emph{The Journal of Chemical Physics} \textbf{2014}, \emph{140},
  044701\relax
\mciteBstWouldAddEndPuncttrue
\mciteSetBstMidEndSepPunct{\mcitedefaultmidpunct}
{\mcitedefaultendpunct}{\mcitedefaultseppunct}\relax
\EndOfBibitem
\bibitem[Miao \latin{et~al.}(2019)Miao, Bellonzi, and Subotnik]{Miao2019}
Miao,~G.; Bellonzi,~N.; Subotnik,~J. An extension of the fewest switches
  surface hopping algorithm to complex Hamiltonians and photophysics in
  magnetic fields: Berry curvature and "magnetic" forces. \emph{Journal of
  Chemical Physics} \textbf{2019}, \emph{150}, 124101\relax
\mciteBstWouldAddEndPuncttrue
\mciteSetBstMidEndSepPunct{\mcitedefaultmidpunct}
{\mcitedefaultendpunct}{\mcitedefaultseppunct}\relax
\EndOfBibitem
\bibitem[Mead(1979)]{Mead1979}
Mead,~C.~A. Molecular dynamics with electronic transitions. \emph{Perspective:
  Nonadiabatic dynamics theory The Journal of Chemical Physics} \textbf{1979},
  \emph{70}, 22--301\relax
\mciteBstWouldAddEndPuncttrue
\mciteSetBstMidEndSepPunct{\mcitedefaultmidpunct}
{\mcitedefaultendpunct}{\mcitedefaultseppunct}\relax
\EndOfBibitem
\bibitem[Berry and Robbins(1993)Berry, and Robbins]{Berry1993}
Berry,~M.~V.; Robbins,~J.~M. Chaotic classical and half-classical adiabatic
  reactions: geometric magnetism and deterministic friction. \emph{Proceedings
  of the Royal Society of London. Series A: Mathematical and Physical Sciences}
  \textbf{1993}, \emph{442}, 659--672\relax
\mciteBstWouldAddEndPuncttrue
\mciteSetBstMidEndSepPunct{\mcitedefaultmidpunct}
{\mcitedefaultendpunct}{\mcitedefaultseppunct}\relax
\EndOfBibitem
\bibitem[Miao \latin{et~al.}(2020)Miao, Bian, Zhou, and Subotnik]{Miao2020}
Miao,~G.; Bian,~X.; Zhou,~Z.; Subotnik,~J. A “backtracking” correction for
  the fewest switches surface hopping algorithm. \emph{The Journal of Chemical
  Physics} \textbf{2020}, \emph{153}, 111101\relax
\mciteBstWouldAddEndPuncttrue
\mciteSetBstMidEndSepPunct{\mcitedefaultmidpunct}
{\mcitedefaultendpunct}{\mcitedefaultseppunct}\relax
\EndOfBibitem
\bibitem[Wu and Subotnik(2021)Wu, and Subotnik]{Wu2021}
Wu,~Y.; Subotnik,~J.~E. Semiclassical description of nuclear dynamics moving
  through complex-valued single avoided crossings of two electronic states.
  \emph{The Journal of Chemical Physics} \textbf{2021}, \emph{154},
  234101\relax
\mciteBstWouldAddEndPuncttrue
\mciteSetBstMidEndSepPunct{\mcitedefaultmidpunct}
{\mcitedefaultendpunct}{\mcitedefaultseppunct}\relax
\EndOfBibitem
\bibitem[Bian \latin{et~al.}(2021)Bian, Wu, Teh, Zhou, Chen, and
  Subotnik]{Bian2021}
Bian,~X.; Wu,~Y.; Teh,~H.-H.; Zhou,~Z.; Chen,~H.-T.; Subotnik,~J.~E. Modeling
  nonadiabatic dynamics with degenerate electronic states, intersystem
  crossing, and spin separation: A key goal for chemical physics. \emph{The
  Journal of Chemical Physics} \textbf{2021}, \emph{154}, 110901\relax
\mciteBstWouldAddEndPuncttrue
\mciteSetBstMidEndSepPunct{\mcitedefaultmidpunct}
{\mcitedefaultendpunct}{\mcitedefaultseppunct}\relax
\EndOfBibitem
\bibitem[Naaman \latin{et~al.}(2019)Naaman, Paltiel, and Waldeck]{Naaman2019}
Naaman,~R.; Paltiel,~Y.; Waldeck,~D.~H. Chiral molecules and the electron spin.
  \emph{Nature Reviews Chemistry 2019 3:4} \textbf{2019}, \emph{3},
  250--260\relax
\mciteBstWouldAddEndPuncttrue
\mciteSetBstMidEndSepPunct{\mcitedefaultmidpunct}
{\mcitedefaultendpunct}{\mcitedefaultseppunct}\relax
\EndOfBibitem
\bibitem[Naaman and Waldeck(2015)Naaman, and Waldeck]{Naaman2015}
Naaman,~R.; Waldeck,~D.~H. The Annual Review of Physical Chemistry is online
  at. \emph{Annu. Rev. Phys. Chem} \textbf{2015}, \emph{66}, 263--281\relax
\mciteBstWouldAddEndPuncttrue
\mciteSetBstMidEndSepPunct{\mcitedefaultmidpunct}
{\mcitedefaultendpunct}{\mcitedefaultseppunct}\relax
\EndOfBibitem
\bibitem[Naaman and Waldeck(2012)Naaman, and Waldeck]{Naaman2012}
Naaman,~R.; Waldeck,~D.~H. Chiral-induced spin selectivity effect.
  \emph{Journal of Physical Chemistry Letters} \textbf{2012}, \emph{3},
  2178--2187\relax
\mciteBstWouldAddEndPuncttrue
\mciteSetBstMidEndSepPunct{\mcitedefaultmidpunct}
{\mcitedefaultendpunct}{\mcitedefaultseppunct}\relax
\EndOfBibitem
\bibitem[Wu \latin{et~al.}(2020)Wu, Miao, and Subotnik]{Wu2020}
Wu,~Y.; Miao,~G.; Subotnik,~J.~E. Chemical Reaction Rates for Systems with
  Spin–Orbit Coupling and an Odd Number of Electrons: Does Berry’s Phase
  Lead to Meaningful Spin-Dependent Nuclear Dynamics for a Two State Crossing?
  \emph{The Journal of Physical Chemistry A} \textbf{2020}, \emph{124},
  7355--7372\relax
\mciteBstWouldAddEndPuncttrue
\mciteSetBstMidEndSepPunct{\mcitedefaultmidpunct}
{\mcitedefaultendpunct}{\mcitedefaultseppunct}\relax
\EndOfBibitem
\bibitem[Wu and Subotnik(2021)Wu, and Subotnik]{Wu2021:CI}
Wu,~Y.; Subotnik,~J.~E. Electronic spin separation induced by nuclear motion
  near conical intersections. \emph{Nature Communications 2021 12:1}
  \textbf{2021}, \emph{12}, 1--7\relax
\mciteBstWouldAddEndPuncttrue
\mciteSetBstMidEndSepPunct{\mcitedefaultmidpunct}
{\mcitedefaultendpunct}{\mcitedefaultseppunct}\relax
\EndOfBibitem
\bibitem[Sakurai and Commins(1995)Sakurai, and Commins]{Sakurai1995}
Sakurai,~J.~J.; Commins,~E.~D. \emph{Modern quantum mechanics, revised
  edition}; American Association of Physics Teachers, 1995\relax
\mciteBstWouldAddEndPuncttrue
\mciteSetBstMidEndSepPunct{\mcitedefaultmidpunct}
{\mcitedefaultendpunct}{\mcitedefaultseppunct}\relax
\EndOfBibitem
\bibitem[Shankar(2012)]{Shankar2012}
Shankar,~R. \emph{Principles of quantum mechanics}; Springer Science \&
  Business Media, 2012\relax
\mciteBstWouldAddEndPuncttrue
\mciteSetBstMidEndSepPunct{\mcitedefaultmidpunct}
{\mcitedefaultendpunct}{\mcitedefaultseppunct}\relax
\EndOfBibitem
\bibitem[foo()]{footnote1}
Consider a scattering process for the system described by Eq.
  \ref{eq:2stateHamiltonian}. The momentum shift effect can be calculated
  through a first-order Born approximation, whereby the scattering amplitude
  between the initial momentum $\bm k_i$ and the final momentum $\bm k_f$ is
  given by $f(\bm k_i - \bm k_f) \propto \int d^2r e^{i (\bm k_i - \bm k_f)
  \cdot \bm r} V(\bm r) = \int dx \sin \theta e^{i (\bm k_i^x - \bm k_f^x)x
  }\int dy e^{i (\bm k_i^y - \bm k_f^y + W) y} = 2\pi \int dx \sin \theta e^{i
  (\bm k_i^x - \bm k_f^x)x } \delta(\bm k_i^y - \bm k_f^y + W)$. The expression
  in the $\delta$ function indicates that the scattering quantum wavepacket
  will accumulate a momentum shift of amount of $W$ in the $y$-direction if the
  wavepacket changes diabat.\relax
\mciteBstWouldAddEndPunctfalse
\mciteSetBstMidEndSepPunct{\mcitedefaultmidpunct}
{}{\mcitedefaultseppunct}\relax
\EndOfBibitem
\bibitem[Wilczek and Zee(1984)Wilczek, and Zee]{Wilczek1984}
Wilczek,~F.; Zee,~A. Appearance of gauge structure in simple dynamical systems.
  \emph{Physical Review Letters} \textbf{1984}, \emph{52}, 2111--2114\relax
\mciteBstWouldAddEndPuncttrue
\mciteSetBstMidEndSepPunct{\mcitedefaultmidpunct}
{\mcitedefaultendpunct}{\mcitedefaultseppunct}\relax
\EndOfBibitem
\bibitem[Mead and Truhlar(1982)Mead, and Truhlar]{Mead1982}
Mead,~C.~A.; Truhlar,~D.~G. Conditions for the definition of a strictly
  diabatic electronic basis for molecular systems. \emph{The Journal of
  Chemical Physics} \textbf{1982}, \emph{77}, 6090--6098\relax
\mciteBstWouldAddEndPuncttrue
\mciteSetBstMidEndSepPunct{\mcitedefaultmidpunct}
{\mcitedefaultendpunct}{\mcitedefaultseppunct}\relax
\EndOfBibitem
\bibitem[Schmidt \latin{et~al.}(2008)Schmidt, Parandekar, and
  Tully]{Schmidt2008}
Schmidt,~J.~R.; Parandekar,~P.~V.; Tully,~J.~C. Mixed quantum-classical
  equilibrium: Surface hopping. \emph{The Journal of Chemical Physics}
  \textbf{2008}, \emph{129}, 044104\relax
\mciteBstWouldAddEndPuncttrue
\mciteSetBstMidEndSepPunct{\mcitedefaultmidpunct}
{\mcitedefaultendpunct}{\mcitedefaultseppunct}\relax
\EndOfBibitem
\bibitem[Kabsch(1976)]{Kabsch1976}
Kabsch,~W. A solution for the best rotation to relate two sets of vectors.
  \emph{Acta Crystallographica Section A} \textbf{1976}, \emph{32},
  922--923\relax
\mciteBstWouldAddEndPuncttrue
\mciteSetBstMidEndSepPunct{\mcitedefaultmidpunct}
{\mcitedefaultendpunct}{\mcitedefaultseppunct}\relax
\EndOfBibitem
\bibitem[Subotnik \latin{et~al.}(2004)Subotnik, Shao, Liang, and
  Head-Gordon]{subotnik:2004:rued}
Subotnik,~J.~E.; Shao,~Y.; Liang,~W.; Head-Gordon,~M. An efficient method for
  calculating maxima of homogeneous functions of orthogonal matrices:
  Applications to localized occupied orbitals. \emph{The Journal of chemical
  physics} \textbf{2004}, \emph{121}, 9220--9229\relax
\mciteBstWouldAddEndPuncttrue
\mciteSetBstMidEndSepPunct{\mcitedefaultmidpunct}
{\mcitedefaultendpunct}{\mcitedefaultseppunct}\relax
\EndOfBibitem
\bibitem[Pacher \latin{et~al.}(1993)Pacher, Cederbaum, and
  K{\"o}ppel]{cederbaum:1993:advchemphys}
Pacher,~T.; Cederbaum,~L.; K{\"o}ppel,~H. Adiabatic and quasidiabatic states in
  a gauge theoretical framework. \emph{Advances in chemical physics}
  \textbf{1993}, \emph{84}, 293--392\relax
\mciteBstWouldAddEndPuncttrue
\mciteSetBstMidEndSepPunct{\mcitedefaultmidpunct}
{\mcitedefaultendpunct}{\mcitedefaultseppunct}\relax
\EndOfBibitem
\bibitem[Subotnik \latin{et~al.}(2013)Subotnik, Ouyang, and
  Landry]{Subotnik2013}
Subotnik,~J.~E.; Ouyang,~W.; Landry,~B.~R. Can we derive Tully's
  surface-hopping algorithm from the semiclassical quantum Liouville equation?
  Almost, but only with decoherence. \emph{The Journal of chemical physics}
  \textbf{2013}, \emph{139}, 211101\relax
\mciteBstWouldAddEndPuncttrue
\mciteSetBstMidEndSepPunct{\mcitedefaultmidpunct}
{\mcitedefaultendpunct}{\mcitedefaultseppunct}\relax
\EndOfBibitem
\bibitem[Kapral and Ciccotti(1999)Kapral, and Ciccotti]{Kapral1999}
Kapral,~R.; Ciccotti,~G. Mixed quantum-classical dynamics. \emph{The Journal of
  chemical physics} \textbf{1999}, \emph{110}, 8919--8929\relax
\mciteBstWouldAddEndPuncttrue
\mciteSetBstMidEndSepPunct{\mcitedefaultmidpunct}
{\mcitedefaultendpunct}{\mcitedefaultseppunct}\relax
\EndOfBibitem
\bibitem[Herman(1984)]{Herman1984}
Herman,~M.~F. Nonadiabatic semiclassical scattering. I. Analysis of generalized
  surface hopping procedures. \emph{The Journal of chemical physics}
  \textbf{1984}, \emph{81}, 754--763\relax
\mciteBstWouldAddEndPuncttrue
\mciteSetBstMidEndSepPunct{\mcitedefaultmidpunct}
{\mcitedefaultendpunct}{\mcitedefaultseppunct}\relax
\EndOfBibitem
\bibitem[Jain \latin{et~al.}(2016)Jain, Alguire, and Subotnik]{Jain2016}
Jain,~A.; Alguire,~E.; Subotnik,~J.~E. An Efficient, Augmented Surface Hopping
  Algorithm That Includes Decoherence for Use in Large-Scale Simulations.
  \emph{Journal of Chemical Theory and Computation} \textbf{2016}, \emph{12},
  5256--5268\relax
\mciteBstWouldAddEndPuncttrue
\mciteSetBstMidEndSepPunct{\mcitedefaultmidpunct}
{\mcitedefaultendpunct}{\mcitedefaultseppunct}\relax
\EndOfBibitem
\bibitem[foo()]{footnote4}
The value of $dt_c$ used for the simulation is 0.1 atomic unit; the value of
  $dt_q$ varies and is calculated on the fly according to Ref.
  \citenum{Jain2016}: \[ dt_q = \begin{cases} dt_c\\ 0.02/\max[\bm T] \\
  0.02\hbar / \max[\bm V - \bm {\bar V}] \end{cases} \] Here, $\max[\bm X]$
  refers to the maximum absolute value of matrix $\bm X$ , $\bm T$ refers to
  the time derivative matrix, $\bm V$ is the potential energy matrix, and $\bm
  {\bar V}$ is the mean value of all adiabatic energies\relax
\mciteBstWouldAddEndPuncttrue
\mciteSetBstMidEndSepPunct{\mcitedefaultmidpunct}
{\mcitedefaultendpunct}{\mcitedefaultseppunct}\relax
\EndOfBibitem
\bibitem[foo()]{footnote2}
Note that this momentum rescaling step is slightly different than the protocol
  in Ref. \citenum{Wu2021}.\relax
\mciteBstWouldAddEndPunctfalse
\mciteSetBstMidEndSepPunct{\mcitedefaultmidpunct}
{}{\mcitedefaultseppunct}\relax
\EndOfBibitem
\bibitem[Hammes‐Schiffer and Tully(1998)Hammes‐Schiffer, and
  Tully]{HammesSchiffer1998}
Hammes‐Schiffer,~S.; Tully,~J.~C. Proton transfer in solution: Molecular
  dynamics with quantum transitions. \emph{The Journal of Chemical Physics}
  \textbf{1998}, \emph{101}, 4657\relax
\mciteBstWouldAddEndPuncttrue
\mciteSetBstMidEndSepPunct{\mcitedefaultmidpunct}
{\mcitedefaultendpunct}{\mcitedefaultseppunct}\relax
\EndOfBibitem
\bibitem[Kosloff and Kosloff(1983)Kosloff, and Kosloff]{Kosloff1983}
Kosloff,~D.; Kosloff,~R. A fourier method solution for the time dependent
  Schrödinger equation as a tool in molecular dynamics. \emph{Journal of
  Computational Physics} \textbf{1983}, \emph{52}, 35--53\relax
\mciteBstWouldAddEndPuncttrue
\mciteSetBstMidEndSepPunct{\mcitedefaultmidpunct}
{\mcitedefaultendpunct}{\mcitedefaultseppunct}\relax
\EndOfBibitem
\bibitem[Mouritsen(2018)]{Mouritsen2018}
Mouritsen,~H. Long-distance navigation and magnetoreception in migratory
  animals. \emph{Nature} \textbf{2018}, \emph{558}, 50--59\relax
\mciteBstWouldAddEndPuncttrue
\mciteSetBstMidEndSepPunct{\mcitedefaultmidpunct}
{\mcitedefaultendpunct}{\mcitedefaultseppunct}\relax
\EndOfBibitem
\bibitem[Kerpal \latin{et~al.}(2019)Kerpal, Richert, Storey, Pillai, Liddell,
  Gust, Mackenzie, Hore, and Timmel]{Kerpal2019}
Kerpal,~C.; Richert,~S.; Storey,~J.~G.; Pillai,~S.; Liddell,~P.~A.; Gust,~D.;
  Mackenzie,~S.~R.; Hore,~P.~J.; Timmel,~C.~R. Chemical compass behaviour at
  microtesla magnetic fields strengthens the radical pair hypothesis of avian
  magnetoreception. \emph{Nature Communications 2019 10:1} \textbf{2019},
  \emph{10}, 1--7\relax
\mciteBstWouldAddEndPuncttrue
\mciteSetBstMidEndSepPunct{\mcitedefaultmidpunct}
{\mcitedefaultendpunct}{\mcitedefaultseppunct}\relax
\EndOfBibitem
\bibitem[Steiner and Ulrichf(1989)Steiner, and Ulrichf]{Steiner1989}
Steiner,~U.~E.; Ulrichf,~T. Magnetic Field Effects in Chemical Kinetics and
  Related Phenomena. \emph{Chem. Rev} \textbf{1989}, \emph{89}, 51\relax
\mciteBstWouldAddEndPuncttrue
\mciteSetBstMidEndSepPunct{\mcitedefaultmidpunct}
{\mcitedefaultendpunct}{\mcitedefaultseppunct}\relax
\EndOfBibitem
\bibitem[Hore \latin{et~al.}(2020)Hore, Ivanov, and Wasielewski]{Hore2020}
Hore,~P.~J.; Ivanov,~K.~L.; Wasielewski,~M.~R. Spin chemistry. \emph{The
  Journal of Chemical Physics} \textbf{2020}, \emph{152}, 120401\relax
\mciteBstWouldAddEndPuncttrue
\mciteSetBstMidEndSepPunct{\mcitedefaultmidpunct}
{\mcitedefaultendpunct}{\mcitedefaultseppunct}\relax
\EndOfBibitem
\bibitem[Higgins \latin{et~al.}(2020)Higgins, Cheisson, Cole, Manor, Carroll,
  and Schelter]{Higgins2020}
Higgins,~R.~F.; Cheisson,~T.; Cole,~B.~E.; Manor,~B.~C.; Carroll,~P.~J.;
  Schelter,~E.~J. Magnetic Field Directed Rare-Earth Separations.
  \emph{Angewandte Chemie International Edition} \textbf{2020}, \emph{59},
  1851--1856\relax
\mciteBstWouldAddEndPuncttrue
\mciteSetBstMidEndSepPunct{\mcitedefaultmidpunct}
{\mcitedefaultendpunct}{\mcitedefaultseppunct}\relax
\EndOfBibitem
\bibitem[foo()]{footnote3}
As an aside, one can also calculate the off-diagonal elements of the Berry
  curvature tensor: ${\bm \Omega}_{ab}^{xy} = \frac {e^{-i(\phi_{a}-\phi_b)}}
  {n} \frac {\partial f(x)} {\partial x} \left(\frac {\partial (\phi_a -
  \phi_b)} {\partial y} - 2 f(x) \frac {\partial \phi_b} {\partial
  y}\right)$\relax
\mciteBstWouldAddEndPuncttrue
\mciteSetBstMidEndSepPunct{\mcitedefaultmidpunct}
{\mcitedefaultendpunct}{\mcitedefaultseppunct}\relax
\EndOfBibitem
\end{mcitethebibliography}
\end{document}